\documentclass[twocolumn,pre,amsmath]{revtex4}

\usepackage[dvipdfmx]{graphicx}

\usepackage{bm}
\usepackage{amsmath}
\usepackage{here}

\begin{document}

\title{
  Softening and residual loss modulus of jammed grains under oscillatory shear in an absorbing state
}

\author{Michio Otsuki}
\email[]{otsuki@me.es.osaka-u.ac.jp}
\affiliation{
  Graduate School of Engineering Science, Osaka University, Toyonaka, Osaka 560-8531, Japan}

\author{Hisao Hayakawa}
\affiliation{Yukawa Institute for Theoretical Physics, Kyoto University, Kitashirakawaoiwake-cho, Sakyo-ku, Kyoto 606-8502, Japan}

\begin{abstract}
  From a theoretical study of the mechanical response of jammed materials comprising frictionless and overdamped particles under oscillatory shear, we find that the material becomes soft, and the loss modulus remains non-zero even in an absorbing state where any irreversible plastic deformation does not exist.    
The trajectories of the particles in this region exhibit hysteresis loops.
  We succeed in clarifying the origin of the softening of the material and the residual loss modulus with the aid of Fourier analysis.
We also clarify the roles of the yielding point in the softening to distinguish the plastic deformation from reversible deformation in the absorbing state. 
\end{abstract}
\date{\today}

\maketitle

{\it Introduction---}
The mechanical response of jammed disordered materials, such as granular materials, foams, emulsions, and colloidal suspensions, garners much attention \cite{Hecke,Behringer}.
For vanishingly small strain, the shear stress $\sigma$ is proportional to the shear strain $\gamma$, which is characterized by the shear modulus satisfying a critical scaling law near the jamming point $\phi_J$ \cite{OHern02,Tighe11,Otsuki17}.
However, the region of the linear response is quite narrow near $\phi_J$ \cite{Coulais,Otsuki14}.
Hence, revealing the nonlinear response is essential for understanding the dynamics of disordered materials.

In crystalline materials, the nonlinear response originates from yielding associated with irreversible plastic deformation.
Yielding also takes place in disordered materials when the strain is sufficiently large \cite{Nagamanasa,Knowlton,Kawasaki16,Leishangthem,Clark,Boschan19}.
The yielding transition attracts much attention among researchers as an example of the reversible-irreversible transition~\cite{Hinrichsen,Henkel,Pine,Corte}.
When plastic deformation causes rearrangements of contact networks,
 the mechanical response becomes nonlinear.
It had been believed that plastic deformation is always necessary for the nonlinear response.
Unlike this expectation, recent studies have revealed that plastic deformation is not always necessary for the nonlinear response \cite{Boschan,Nakayama,Kawasaki20,Bohy,Ishima}.
Under steady shear, $\sigma$ becomes hypoelastic before the yielding \cite{Boschan,Kawasaki20}, and the storage modulus in the steady state after applying a sufficient number of cyclic shears decreases as the strain amplitude increases without any irreversible plastic deformation \cite{Bohy}.
The decrease of the storage modulus is called softening.

It is known that plastic deformation causes dissipation characterized by the loss modulus \cite{Bohy,Ishima}.
It is natural that the loss modulus disappears in quasi-static strains without any plastic deformation.
However, we need careful check of this naive picture, because the loss modulus might be related to the softening observed without any plastic deformation.

The mechanical response should be related to the motion of particles constituting the disordered materials.
This suggests that the trajectories of particles provide information on the softening of the materials.
Several studies have reported that the trajectories of dense particles form closed loops under oscillatory shear below the yielding point associated with reversible contact changes where there are some cyclic open and closed contacts between particles~\cite{Lundberg,Schreck,Keim13,Keim14,Regev13,Regev15,Priezjev,Lavrentovich,Nagasawa,Das,Deen,Khirallah}.
The formation of closed loops means that the system is reduced to an absorbing state after some time has passed. 
A previous study numerically showed that the softening in the absorbing state becomes significant when there are closed loops associated with many contact changes. 
However, the quantitative relationship remains unclear \cite{Bohy}.

In this study, we numerically investigate jammed materials comprising $N$ frictionless and overdamped particles under oscillatory shear to clarify the origin of the softening.
For this purpose, we focus on the roles of the trajectories to clarify the relationship between the softening in the absorbing state and the softening in the plastic regime.
We find that the shear modulus exhibits softening, and the loss modulus remains non-zero even in the absorbing state below the yielding point.
The trajectory of a test particle forms a nontrivial loop in this region.
With the aid of Fourier analysis, we investigate the geometric structure of the trajectories and reveal the role of Fourier components for the storage and loss moduli.
We also present the theoretical expressions for the storage and loss moduli, whose quantitative validities are numerically confirmed.

{\it Setup---}
Let us consider a jammed two-dimensional system consisting of frictionless particles under oscillatory shear.
The particles are driven by the overdamped equation with Stokes' drag under Lees--Edwards boundary conditions \cite{Evans}, where the equation of motion is given by
\begin{equation}
  \zeta \left \{ \frac{d}{dt} {\boldsymbol r}_i - \dot \gamma(t) y_i \boldsymbol e_x \right \} 
  = - \sum_{j\neq i} \frac{\partial}{\partial \boldsymbol r_i} U(r_{ij})
\end{equation}
with the position ${\boldsymbol r}_i = (x_i, y_i)$ of particle $i$.
Here, $\zeta$ and $\dot \gamma (t)$ are the drag coefficient and strain rate, respectively.
The interaction potential $U(r_{ij})$ is assumed to be 
\begin{equation}
  U(r_{ij}) = \frac{k}{2} (d_{ij} - r_{ij})^2 \Theta(d_{ij} - r_{ij}),
  \label{eq:U}
\end{equation}
where $\Theta(x)$, $k$, $d_{ij}$, and $r_{ij}=|\boldsymbol r_{ij}|=|\boldsymbol r_i - \boldsymbol r_i|$ are the Heaviside step function satisfying $\Theta(x)=1$ for $x\ge 0$ and $\Theta(x)=0$ otherwise, the spring constant, the average diameter of particles $i$ and $j$, and the distance between particles $i$ and $j$, respectively.
The system is bidisperse and consists of an equal number of particles with diameters $d_0$ and $d_0/1.4$.
We have verified that particles with inertia and damping at contact, which corresponds to the model in Ref. \cite{Bohy}, exhibit almost identical behavior in our system \cite{Supple}.

We prepare the initial state with a given packing fraction $\phi$ by slowly compressing the system from a state below the jamming point $\phi_{\rm J} \simeq 0.841$ \cite{Otsuki17}.
The oscillatory shear strain is applied for $n_c$ cycles as
\begin{equation}
  \gamma(\theta) = \gamma_0 \sin \theta
  \label{Eq:shear}
\end{equation}
with the phase $\theta = \omega t$, where $\gamma_0$ and $\omega$ are the strain amplitude and angular frequency, respectively.
Note that the shear rate satisfies $\dot{\gamma}(t)=(d\theta/dt) (d/d\theta)\gamma(\theta)$.
In the last cycle, we measure the storage and loss moduli $G'$ and $G''$, respectively, given by \cite{Doi}
\begin{eqnarray}
  \label{eq:storage modulus}
  G' & = & \frac{1}{\pi} \int_0^{2\pi} \ d\theta \  \frac{\left \langle \sigma(\theta) \right \rangle \sin \theta}{\gamma_0}, \\
  \label{eq:loss modulus}
  G'' & = & \frac{1}{\pi} \int_0^{2\pi} \ d\theta \ \frac{\left \langle \sigma(\theta) \right \rangle \cos \theta}{\gamma_0}, 
\end{eqnarray}
with shear stress
\begin{equation}
  \sigma  = \frac{1}{L^2} \sum_{i} \sum_{j>i} \frac{x_{ij}y_{ij}}{r_{ij}}U'(r_{ij}),
  \label{eq:stress}
\end{equation}
where $x_{ij}= x_i - x_j$, $y_{ij}= y_i - y_j$, $\langle \cdot \rangle$ represents the ensemble average, and $L$ is the linear system size.
See Ref. \cite{Supple} for the stress-strain curves in our system.
We have verified that $G'$ and $G''$ are independent of $N$ and $n_c$ for $N \ge 1000$ and $n_c \ge 20$ \cite{Supple}. 
We use $N=1000$ and $n_c=20$ in our numerical analysis.
We adopt the Euler method using the time step $\Delta t = 0.05 \tau_0$ with $\tau_0 =\zeta/k$.

\begin{figure}[htbp]
  \begin{center}
    \begin{tabular}{c}
      \begin{minipage}{0.5\hsize}
      \begin{center}
        \includegraphics[width=1.0\linewidth]{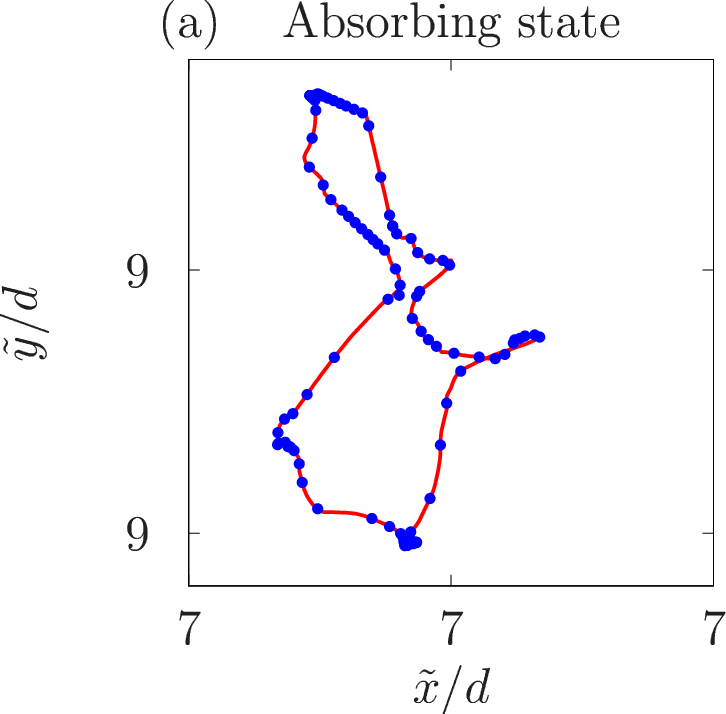}
      \end{center}
      \end{minipage}
      \begin{minipage}{0.50\hsize}
      \begin{center}
        \includegraphics[width=1.0\linewidth]{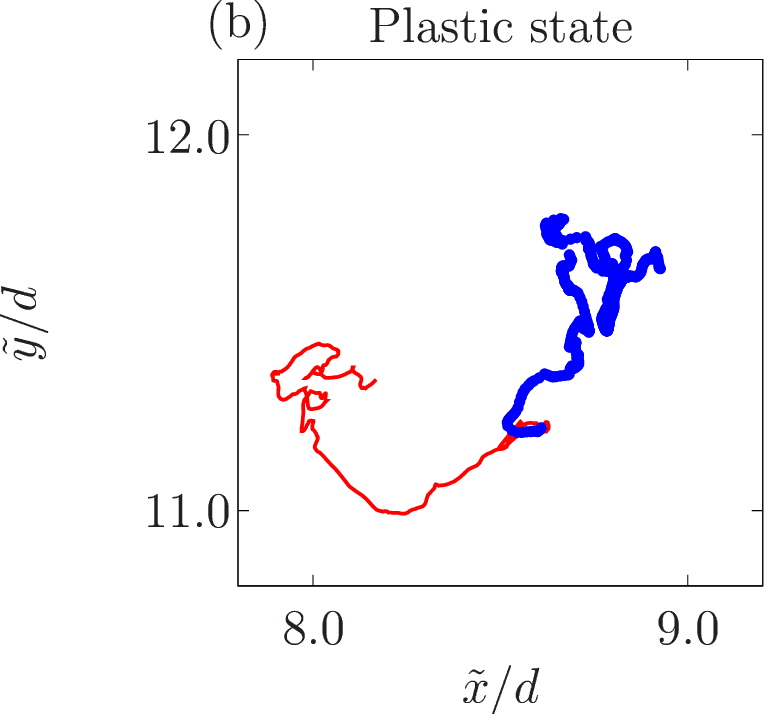}
      \end{center}
      \end{minipage}
    \end{tabular}
\caption{
  Non-affine particle trajectories in the last two cycles for $\gamma_0 = 0.02$ (a) and $0.1$ (b) with $\omega=10^{-4}\tau_0^{-1}$ and $\phi=0.87$, which corresponds to $\phi - \phi_J=0.029$.
    The circles represent the trajectory in the last cycle. The line represents the trajectory in the second to the last cycle.
}
\label{trjna}
  \end{center}
\end{figure}

{\it Closed Trajectories---}
As the number of cycles increases, the system reaches a statistically steady state through a transient regime as shown in Ref. \cite{Supple}.
Figure \ref{trjna} displays typical non-affine trajectories of a particle
\begin{equation}
\tilde {\boldsymbol r}_i(\theta)
= \boldsymbol r_i(\theta)
  - \gamma(\theta) y_i(\theta) \boldsymbol e_x 
  \label{eq:deconposition}
\end{equation}
in the last two cycles with $\phi=0.87$ and $\omega=10^{-4}\tau_0^{-1}$ in the steady state.
In Fig.\ref{trjna}(a) ($\gamma_0 = 0.02$), the trajectories are closed, and the particle returns to its original position after every cycle. 
This indicates that irreversible plastic deformation does not occur, at least in the last two cycles.
The closed trajectories form nontrivial loops, which differ from ellipses or lines observed for small $\gamma_0$ as shown in Ref. \cite{Supple}.
In Fig. \ref{trjna}(b) ($\gamma_0 = 0.1$), the particle moves away from its original positions after a cycle, as a characteristic behavior of plastic deformation.
Here, we define the absorbing state where the displacement of each particle after several cycles is smaller than $d_c = 10^{-4} d_0$ in the statistically steady state.
We also define the plastic state where the displacement after several cycles exceeds $d_c$.
It should be noted that some rare samples exhibit trajectories where particles return to their original positions after more than one cycle \cite{Regev13,Regev15,Lavrentovich,Nagasawa,Khirallah}. 
However, our theoretical results shown below are unchanged even if such samples exist \cite{Supple}.

\begin{figure}[htbp]
\includegraphics[width=0.9\linewidth]{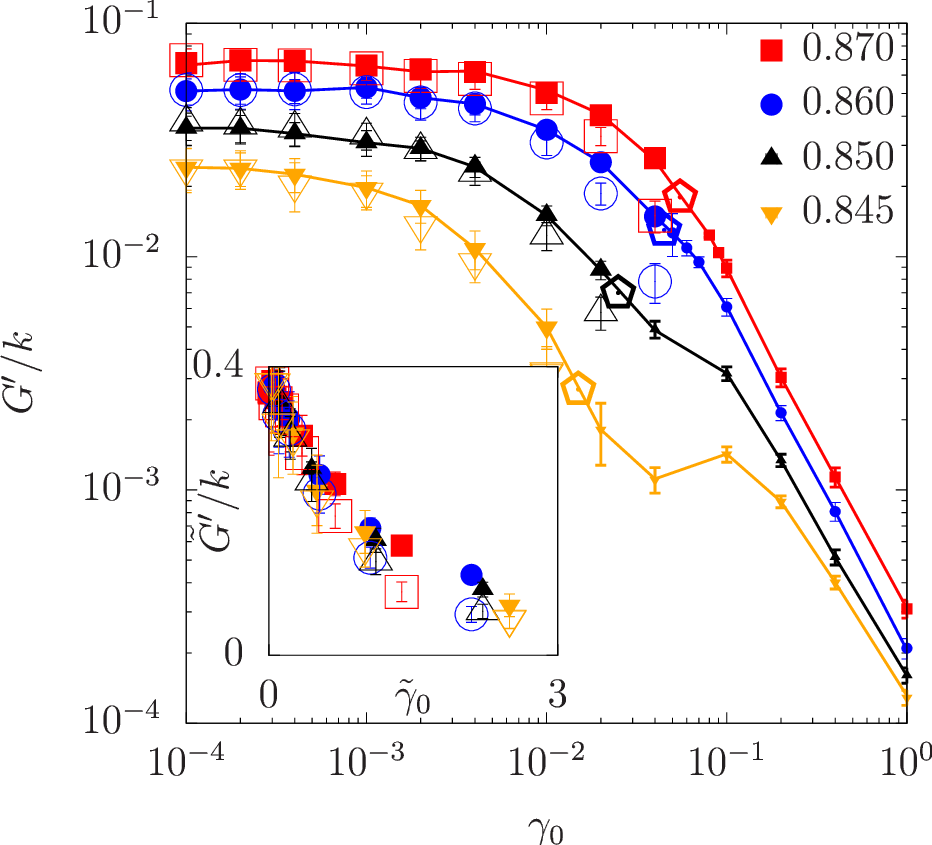}
  \caption{ Storage modulus $G'$ obtained in our simulation (filled symbols) against $\gamma_0$ for $\omega = 10^{-4}\tau_0^{-1}$ with $\phi=0.870, 0.860, 0.850,$ and $0.845$, which corresponds to $\phi - \phi_J=0.029, 0.019, 0.009$, and $0.004$, respectively.
  The legends represent the packing fraction $\phi$.
  The data in the absorbing (plastic) state obtained in our simulation are shown in larger (smaller) filled symbols.
  The open pentagons represent the yielding strain amplitude $\gamma_c$, while other open symbols represent the theoretical expression using $G'_{\rm T}$ in Eq. \eqref{Eq:Gp1}.
  (Inset) Scaled storage modulus $\tilde G' = G'/\sqrt{\phi - \phi_J}$ obtained in our simulation (filled symbols) and its theoretical expression using $G'_{\rm T}$ (open symbols) in Eq. \eqref{Eq:Gp1} against scaled strain amplitude $\tilde \gamma_0 = \gamma_0/(\phi - \phi_J)$ in the absorbing state.
}
\label{Gp}
\end{figure}

{\it Shear Modulus---}
We plot the storage modulus $G'$ against the strain amplitude $\gamma_0$ for $\omega = 10^{-4}\tau_0^{-1}$ with $\phi=0.870, 0.860, 0.850,$ and $0.845$ in Fig. \ref{Gp}.
The yielding points to distinguish the absorbing state from the plastic state for various $\phi$
are shown by open pentagons \cite{Supple}.
The storage modulus $G'$ decreases as $\gamma_0$ increases, but the yielding point is not identical to the point where $G'$ starts to decrease.
We call the decrease for $\gamma_0 < \gamma_c$, the yielding strain amplitude, the softening in the absorbing state (SAS).
We also call the decrease for $\gamma_0 > \gamma_c$ the softening in the plastic state (SPS).
It is remarkable that SAS is continuously connected to SPS, while a shoulder in $G'$ appears in SPS for $0.04 \le \gamma_0 \le 0.1$ with $\phi=0.845$.
In the inset of Fig. \ref{Gp}, we demonstrate that $G'$ and $\gamma_0$ can be scaled by $\sqrt{\phi - \phi_J}$ and $\phi - \phi_J$, respectively, as indicated in Refs. \cite{OHern02, Bohy}.
We have confirmed that $G'$ is independent of $\omega$ for $\omega \le 10^{-3}\tau_0^{-1}$.

\begin{figure}[htbp]
\includegraphics[width=0.9\linewidth]{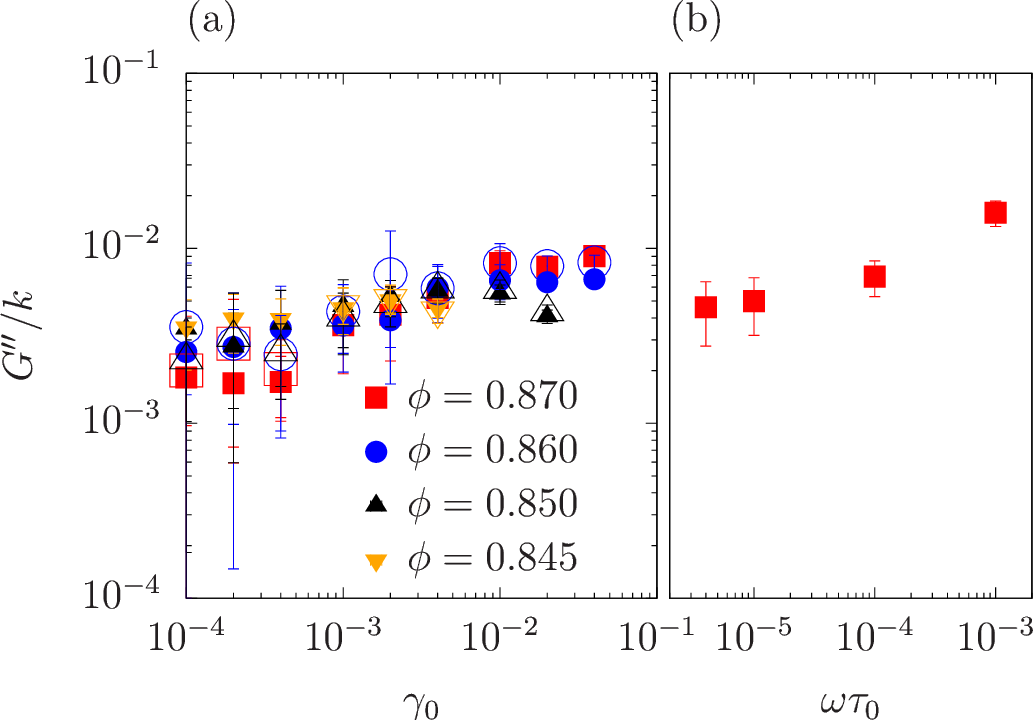}
  \caption{(a) Loss modulus $G''$ in the absorbing state obtained in our simulation (filled symbols) and its theoretical expression $G''_{\rm T}$ (open symbols) in Eq. \eqref{Eq:Gpp1} against $\gamma_0$ for $\omega = 10^{-4}\tau_0^{-1}$ with $\phi=0.870, 0.860, 0.850,$ and $0.845$, which corresponds to $\phi - \phi_J=0.029, 0.019, 0.009$, and $0.004$, respectively.
  (b) Loss modulus $G''$ against $\omega \tau_0$ for $\phi=0.87$ with $\gamma_0=0.01$.
}
\label{Gpp}
\end{figure}

Figure \ref{Gpp}(a) displays the loss modulus $G''$ in the absorbing state against $\gamma_0$ for $\omega = 10^{-4}\tau_0^{-1}$ with $\phi=0.870, 0.860, 0.850,$ and $0.845$, in which $G''$ does not strongly depend on $\phi$ and $\gamma_0$.
See Ref. \cite{Supple} for $G''$ in the plastic state.
In Fig. \ref{Gpp}(b), we plot the loss modulus $G''$ in the absorbing state against $\omega$ for $\phi=0.87$ with $\gamma=0.01$.
Remarkably, $G''$ in Fig. \ref{Gpp}(b) seems to converge to a non-zero value in the limit $\omega \to 0$, which contrasts with the behavior of the Kelvin--Voigt  model (i.e., $G'' \propto \omega$ \cite{Meyers}).
This behavior indicates that dissipation remains even in the quasi-static limit in the absorbing state.
Note that $G'' \propto \omega$ is recovered when we adopt a sufficiently small $\gamma_0$ \cite{Supple}.

\begin{figure}[htbp]
\includegraphics[width=0.8\linewidth]{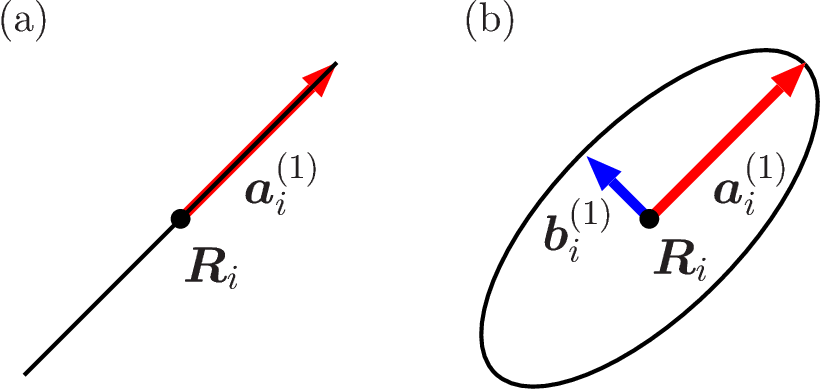}
  \caption{
    Schematics of the non-affine trajectory when only $\boldsymbol a_i^{(1)}$ is non-zero (a) and only $\boldsymbol a_i^{(1)}$ and $\boldsymbol b_i^{(1)}$ are non-zero (b).
}
\label{trajectory}
\end{figure}

{\it Fourier Analysis---}
In the absorbing state, the non-affine trajectory $\tilde {\boldsymbol r}_i(\theta)$ of particle $i$ can be expressed in a Fourier series as 
\begin{equation}
  \tilde {\boldsymbol r}_i(\theta) =   \boldsymbol R_i +
  \sum_{n=1}^\infty \left ( \boldsymbol a_i^{(n)} \sin n \theta
  + \boldsymbol b_i^{(n)} \cos n \theta \right )
  \label{eq:Fourier}
\end{equation}
with the center of the trajectory
\begin{eqnarray}
  \label{eq:R}
  \boldsymbol R_i  & = & (X_i, Y_i) = \frac{1}{2\pi}
  \int_0^{2\pi} \ d\theta \ \tilde{\boldsymbol r}_i(\theta),
\end{eqnarray}
and the Fourier coefficients
\begin{eqnarray}
  \label{eq:a}
  \boldsymbol a_i^{(n)}  & = & \frac{1}{\pi}
  \int_0^{2\pi} \ d\theta \ \sin n\theta \ \tilde{\boldsymbol r}_i(\theta), \\
  \label{eq:b}
  \boldsymbol b_i^{(n)}  & = & \frac{1}{\pi}
  \int_0^{2\pi} \ d\theta \ \cos n\theta \ \tilde{\boldsymbol r}_i(\theta).
\end{eqnarray}
If $\boldsymbol a_i^{(n)} = \boldsymbol b_i^{(n)} = \boldsymbol 0$ for all $n$, the particle motion is affine.
When only $\boldsymbol a_i^{(1)}$ is non-zero, the non-affine trajectory is a straight line, as shown in Fig. \ref{trajectory}(a).
In contrast, the trajectory exhibits an ellipse when $\boldsymbol b_i^{(1)}$ is also non-zero, as shown in Fig. \ref{trajectory}(b).
A nontrivial trajectory, as shown in Fig. \ref{trjna}(a), contains modes with $n\ge 2$.
See Ref. \cite{Supple} for the relationship between the trajectories and the Fourier coefficients.

\begin{figure}[htbp]
  \begin{center}
    \begin{tabular}{c}
      \begin{minipage}{0.5\hsize}
      \begin{center}
        \includegraphics[width=1.0\linewidth]{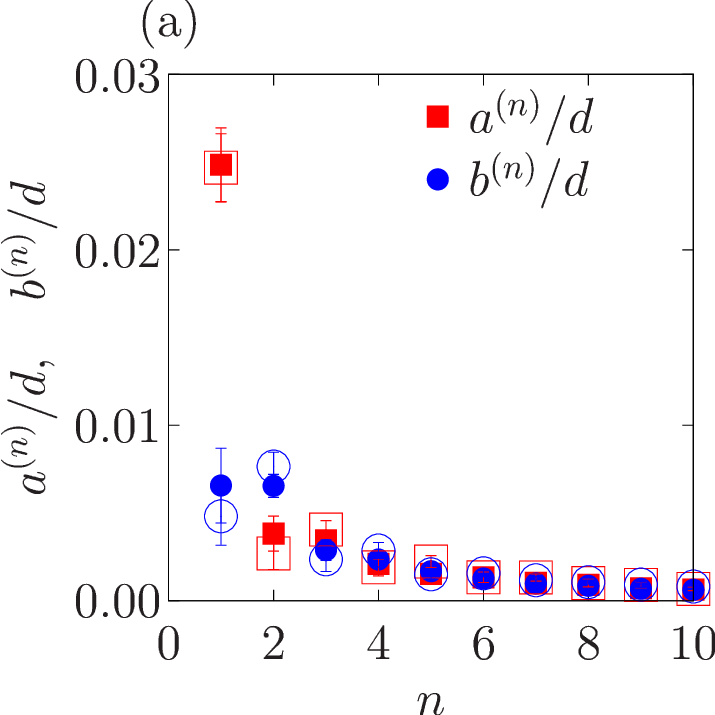}
      \end{center}
      \end{minipage}
      \begin{minipage}{0.5\hsize}
      \begin{center}
        \includegraphics[width=0.9\linewidth]{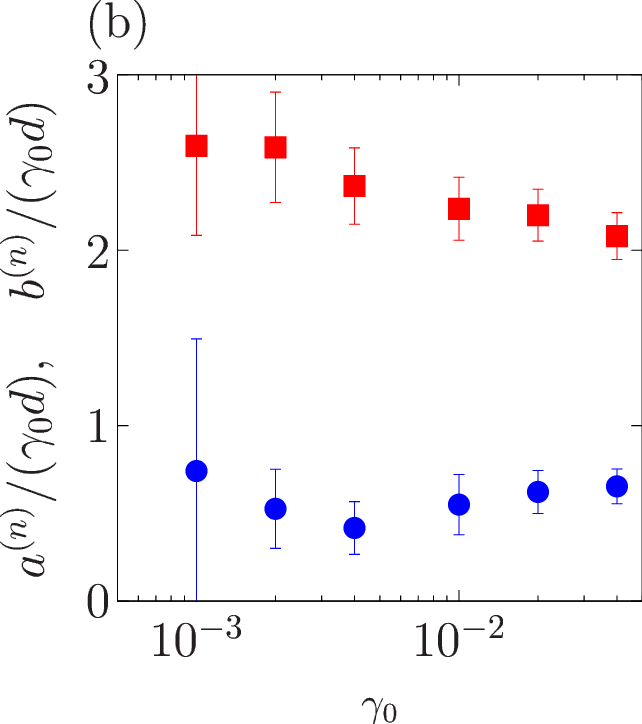}
      \end{center}
      \end{minipage}
    \end{tabular}
    \caption{(a) Magnitudes of Fourier coefficients $a^{(n)}$ and $b^{(n)}$ against $n$ for $\phi=0.87$ and $\gamma_0 = 0.02$ with $\omega \tau_0 = 10^{-4}$ (filled symbols) and $10^{-5}$ (open symbols).
  (b) Magnitudes of the Fourier coefficients $a^{(n)}$ and $b^{(n)}$ normalized by $\gamma_0$ against $\gamma_0$ for $\phi=0.87$ and $\omega \tau_0 = 10^{-4}$ with $n=1$.
$\phi = 0.87$ corresponds to $\phi - \phi_J=0.029$.
}
\label{Fourier}
  \end{center}
\end{figure}

In. Fig. \ref{Fourier}(a), we plot the magnitudes of the Fourier components
\begin{equation}
  a^{(n)}  =  \sum_i \left \langle \left | \boldsymbol a_i^{(n)} \right | \right \rangle/N , \ \
  b^{(n)}  =   \sum_i \left \langle \left | \boldsymbol b_i^{(n)} \right | \right \rangle / N 
\end{equation}
obtained from our numerical data using Eqs. \eqref{eq:a} and \eqref{eq:b}
against $n$ for $\phi=0.87$ and $\gamma_0 = 0.01$ with $\omega \tau_0 = 10^{-4}$ and $10^{-5}$.
The Fourier components do not strongly depend on $\omega$, which indicates that the nontrivial loops do not disappear in the limit $\omega \to 0$.
For different $\phi > \phi_J$ and $\gamma_0 \ge 10^{-3}$, we have confirmed that $a^{(1)}$ is always the largest \cite{Comment}, the other modes are non-zero to make loops with non-zero areas, and the Fourier components are independent of $\omega$.
In Fig. \ref{Fourier}(b), we plot
$a^{(n)}/\gamma_0$ and $b^{(n)}/\gamma_0$ against $\gamma_0$ for $\phi=0.87$ and $\omega \tau_0  = 10^{-4}$ with $n=1$, where
$a^{(n)}/\gamma_0$ and $b^{(n)}/\gamma_0$ are almost independent of $\gamma_0$.
This behavior is consistent with that for the number of contact changes \cite{Supple}.

{\it Theoretical Analysis---}
Now, let us reproduce the numerical results by a simple analytic calculation.
Substituting Eq. \eqref{eq:Fourier} into Eq. \eqref{eq:deconposition}, ${\boldsymbol r}_{ij}(\theta)$ is given by
\begin{eqnarray}
  {\boldsymbol r}_{ij}(\theta) & = &  \boldsymbol R_{ij} + \gamma_0 Y_{ij} \sin \theta \boldsymbol e_x  \nonumber \\
  &  & +
  \sum_{n=1}^\infty \left ( \boldsymbol a_{ij}^{(n)} \sin n \theta
  + \boldsymbol b_{ij}^{(n)} \cos n \theta \right )
  \label{eq:rapp}
\end{eqnarray}
Here, we define $\boldsymbol a^{(n)}_{ij} =  \boldsymbol a^{(n)}_{i}-\boldsymbol a^{(n)}_{j}$, $\boldsymbol b^{(n)}_{ij} =  \boldsymbol b^{(n)}_{i}-\boldsymbol b^{(n)}_{j}$, and $\boldsymbol R_{ij} = (X_{ij},Y_{ij}) = \boldsymbol R_{i}-\boldsymbol R_{j}$.
Substituting Eq. \eqref{eq:rapp} into Eq. \eqref{eq:storage modulus} with Eq. \eqref{eq:stress} and neglecting the terms of $O(\gamma_0)$, we obtain the expression $G'_{\rm T}$ of the storage modulus in SAS as \cite{Supple}
\begin{eqnarray}
  G'_{\rm T} & = & 
  - \frac{1}{L^2}\sum_{i,j} \left \langle 
\frac{X_{ij}^2 Y_{ij}^2 }{R_{ij}}\Psi'(R_{ij}) 
  \right \rangle \nonumber \\
 &  & 
  - \frac{1}{L^2}\sum_{i,j} \left \langle
Y_{ij}^2 \Psi(R_{ij})  
   \right \rangle \nonumber \\
  & & - \frac{1}{L^2}\sum_{i,j} \left \langle  \left ( \frac{a_{ij,x}^{(1)}}{\gamma_0}Y_{ij}  + X_{ij} \frac{a_{ij,y}^{(1)}}{\gamma_0} \right ) \Psi(R_{ij})  \right \rangle \nonumber \\
  & & - \frac{1}{L^2}\sum_{i,j} \left \langle X_{ij}Y_{ij}\Psi'(R_{ij})
   \frac{\boldsymbol R_{ij} \cdot \boldsymbol a_{ij}^{(1)}}{\gamma_0 R_{ij}} \right \rangle,
   \label{Eq:Gp1}
\end{eqnarray}
where $\Psi(r) = -U'(r)/r$.
Here, we have assumed $|a_i^{(n)}| \sim |b_i^{(n)}| \sim \gamma_0$ and $\gamma_0 \ll 1$.
In the expression of Eq. \eqref{Eq:Gp1}, only the first harmonic contribution from $\bm{a}_i^{(1)}$ can survive because of Eq. \eqref{eq:storage modulus}.
Note that $\boldsymbol R_{i}$ and $\boldsymbol a^{(1)}_i$ cannot be determined within the theory but are determined by our simulation data.
In Fig. \ref{Gp}, we plot the theoretical prediction $G'_{\rm T}$ as open symbols. 
The theoretical prediction quantitatively reproduces the numerical results except for large $\gamma_0$, which is out of the scope of our theory.
The first and second terms on the right-hand side (RHS) of Eq. \eqref{Eq:Gp1} represent the contributions from the affine transformation depending only on $\boldsymbol R_{i}$, while the third and fourth terms including $\boldsymbol a^{(1)}_{ij}/\gamma_0$ indicate the contributions from the non-affine trajectories.
As shown in Ref. \cite{Supple}, the contributions from the non-affine trajectories are almost independent of $\gamma_0$, which is consistent with the behavior of $a^{(1)}/\gamma_0$ shown in Fig. \ref{Fourier}(b).
Numerical evaluation in Ref. \cite{Supple} reveals that SAS is dominated by the first term on RHS of Eq. \eqref{Eq:Gp1} through the change of $\boldsymbol R_i$.
The center of the non-affine trajectories $\boldsymbol R_i$ is changed by the rearrangement of the configuration during the transient to the absorbing state, which is consistent with the memory formation of dense particles during oscillatory shear \cite{Fiocco, Paulsen, Adhikari}.

The theoretical expression $G''_{\rm T}$ of the loss modulus in SAS is given by \cite{Supple} 
\begin{eqnarray}
  G_{\rm T}''  & = & 
   - \frac{1}{L^2}\sum_{i,j}\left \langle \left ( \frac{b_{ij,x}^{(1)}}{\gamma_0} Y_{ij} + X_{ij}\frac{b_{ij,y}^{(1)}}{\gamma_0} \right )  \Psi(R_{ij})  \right \rangle \nonumber \\
   & & 
  - \frac{1}{L^2}\sum_{i,j}\left \langle X_{ij}Y_{ij}\Psi'(R_{ij}) R_{ij}
   \frac{\boldsymbol R_{ij}\cdot  \boldsymbol b_{ij}^{(1)}}{\gamma_0 R_{ij}^2} \right \rangle,
   \label{Eq:Gpp1}
\end{eqnarray}
where we have used the same assumption to obtain Eq. \eqref{Eq:Gp1}.
Similar to the case of $G_T'$, only the contribution of the first harmonics $\bm{b}_i^{(1)}$ in the expression of Eq. \eqref{eq:Fourier} can survive because of Eq. \eqref{eq:loss modulus}.
Note that $\bm{b}_i^{(1)}$ cannot be determined within the theory but is evaluated by the simulation data.
The loss modulus depends only on the non-affine contribution including ${\boldsymbol b}_i^{(1)}$.  The amplitude $b^{(1)}$ remains non-zero in the limit $\omega \to 0$, which leads to the residual loss modulus as in Fig. \ref{Gpp}(b).
We plot the theoretical expression $G''_{\rm T}$ using the open symbols in Fig. \ref{Gpp}(a). 
$G''_{\rm T}$ also reproduces the numerical results except for large $\gamma_0$.
Thus, our theory reveals the quantitative relationship between the loss modulus and closed trajectories, which was suggested in Ref. \cite{Keim14}.

{\it Conclusion---}
We numerically studied the mechanical response of jammed materials consisting of frictionless and overdamped particles under oscillatory shear.
The shear modulus exhibits SAS and the residual loss modulus exists in the quasi-static limit in the absorbing state.
Through Fourier analysis of the closed trajectories, the theoretical expressions for the storage and loss moduli quantitatively agree with the numerical results.

Reference \cite{Tighe11} reported that the loss modulus vanishes in the absorbing jammed states in the limit $\omega \to 0$, which is inconsistent with our result.
It is noteworthy that Ref. \cite{Tighe11} did not consider any transient state associated with contact changes before the system reaches the absorbing state.
Since the loss modulus is expected to be given by the generalized Green-Kubo formula \cite{Chong,Hayakawa}, the origin of the residual loss modulus might be plastic events in the transient dynamics.

Recent studies of large amplitude oscillatory shear (LAOS) reveal that there are contributions from higher harmonics in the mechanical response of nonlinear viscoelastic materials \cite{Wagner, Hyun}.
We calculate nonlinear viscoelastic moduli $G'_n$ and $G''_n$ with $n \ge 2$ and confirm that such higher order moduli are negligible in our system as shown in Ref. \cite{Supple}.

In this Letter, we focus only on the nonlinear response of disordered frictionless particles.
However, even frictional grains and exhibit SAS depending on the friction coefficient \cite{Otsuki21}.
Therefore, an extension of our theory to these systems will be our future work.

\begin{acknowledgments}
The authors thank K. Saitoh, D. Ishima, T. Kawasaki, K. Miyazaki, and K. Takeuchi for fruitful discussions.
This work was supported by JSPS KAKENHI Grants No. JP16H04025 and No. JP19K03670 and ISHIZUE 2020 of the Kyoto University Research Development Program.
\end{acknowledgments}

\clearpage

\setcounter{equation}{0}
\setcounter{figure}{0}

\renewcommand{\theequation}{S\arabic{equation}}
\renewcommand{\thefigure}{S\arabic{figure}}

\begin{center}
\textbf{\Large Supplemental Material: }
\end{center}

This Supplemental Material provides some details that are not written in the main text. 
The results for underdamped frictionless granular particles without background friction are presented in Sec. \ref{Sec:GR}.
In Sec. \ref{Sec:Nc}, we show the dependence of $G'$ and $G''$ on the number of particles $N$ and the number of cycles $n_c$.
In Sec. \ref{Sec:yield}, we present the time evolution of the displacements of particles before reaching the absorbing state and the evaluation of the yielding strain amplitude $\gamma_c$.
In Sec. \ref{Sec:stress}, we illustrates the time evolutions of the stress-strain curves in the absorbing and plastic states.
In Sec. \ref{Sec:LoopDep}, we show how particle trajectories depend on $\gamma_0$ and $\omega$.
In Sec. \ref{Sec:double}, we show that trajectories in the absorbing state with longer periods do not affect our theoretical results based on the absorbing trajectories whose periods are identical to the period of the external oscillation.
In Sec. \ref{Sec:Loss}, we present the loss modulus in the absorbing and plastic states.
In Sec. \ref{Sec:KV}, we demonstrate how the naive result of the Kelvin--Voigt model can be recovered for sufficiently small strain amplitude.
In Sec. \ref{loop}, we illustrate the relation between the Fourier coefficients and the shape of particle trajectories. 
In Sec. \ref{Sec:CC}, we show the number of contact changes during the last cycle in the absorbing state.
In Sec. \ref{Sec:theory}, we derive Eqs. (14) and (15) in the main text.
In Sec. \ref{Sec:Components}, we decompose the storage and loss moduli into several components, and clarify what components are dominant contributions for the storage and loss moduli.
In Sec. \ref{Sec:HH}, we show the nonlinear viscoelastic moduli in our system to clarify the roles of higher harmonics.

\section{Underdamped granular particles}
\label{Sec:GR}

In this section, we show that our results are qualitatively unchanged in underdamped frictionless granular particles without background friction.
Here, we use the SLLOD equation given by [34] 
\begin{eqnarray}
  \frac{d}{dt} {\boldsymbol r}_i & = & \dot \gamma(t) y_i \boldsymbol e_x + \boldsymbol p_i, \\
   \frac{d}{dt} {\boldsymbol p}_i  & = & - \dot \gamma(t) p_{i,y} \boldsymbol e_x + \boldsymbol F_i
\end{eqnarray}
under the Lees--Edwards boundary condition,
where
$\bm{p}_i=m(\dot{\bm{r}}_i-\dot\gamma(t)y_i) \bm{e}_x$ and
\begin{equation}
\boldsymbol F_i = 
- \sum_{j\neq i} \frac{\partial}{\partial \boldsymbol r_i} U(r_{ij})
  - \sum_{j\neq i} \eta v^{\rm (n)}_{ij} \Theta(d_{ij} - r_{ij}) \frac{\boldsymbol r_{ij}}{r_{ij}}
\end{equation}
with mass $m$, the interaction potential $U(r_{ij})$ given by Eq. (2), the viscous constant $\eta$, and the normal velocity 
\begin{eqnarray}
v^{\rm (n)}_{ij} =  
  \left \{ \frac{d}{dt} \boldsymbol r_i - \frac{d}{dt} \boldsymbol r_j \right \}\cdot \frac{\boldsymbol r_{ij}}{r_{ij}}.
\end{eqnarray}
We adopt $\eta = \sqrt{mk}$.
This model corresponds to frictionless granular particles with the restitution coefficient $e = 0.043$.
We adopt the leapfrog algorithm using the time step $\Delta t = 0.05 t_0$
with the characteristic time with $t_0 = \sqrt{m/k}$.

\begin{figure}[htbp]
  \begin{center}
        \includegraphics[width=0.5\linewidth]{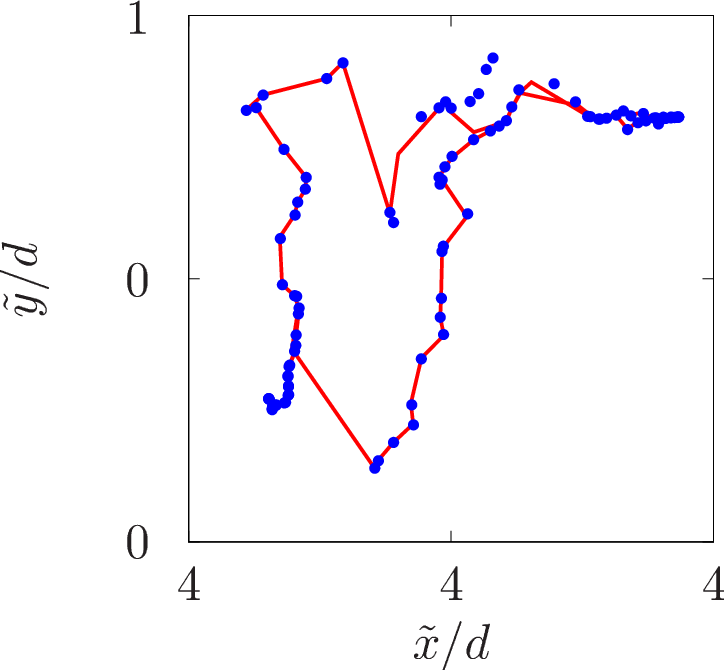}
\caption{
Non-affine trajectories in the last two cycles for underdamped particles with $\gamma_0 = 0.02$, $\phi=0.87$, and $\omega=10^{-4}\tau_0^{-1}$ in the absorbing state.
The circles represent the trajectory in the last cycle.
The line represents the trajectory in the second to the last cycle.
}
\label{trjna_GR}
  \end{center}
\end{figure}

In Fig. \ref{trjna_GR}, we plot non-affine trajectories in the last two cycles for a particle with $\gamma_0 = 0.02$, $\phi=0.87$, and $\omega=10^{-4}\tau_0^{-1}$ in the absorbing state.
The trajectories are closed, but there is a region where the position of the particle depends on the number of cycles $n_c$.
The result of Fig. \ref{trjna_GR} is a typical one from the inertia effect in the underdamped system.

\begin{figure}[htbp]
\includegraphics[width=0.8\linewidth]{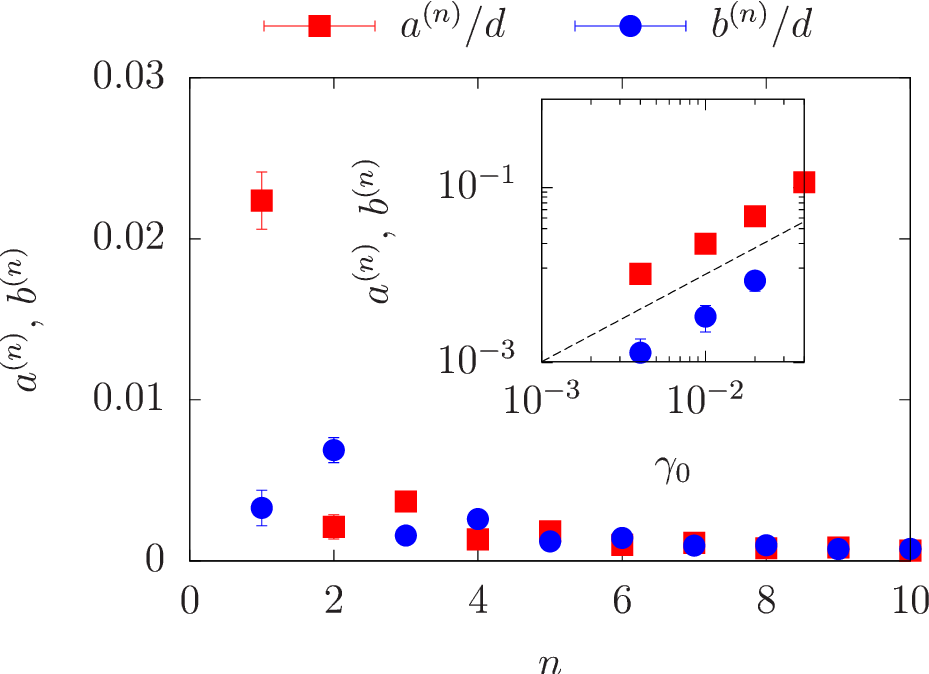}
  \caption{Magnitudes of Fourier coefficients $a^{(n)}$ and $b^{(n)}$ of the underdamped particles against $n$ for $\phi=0.87$ and $\gamma_0 = 0.02$ with $\omega \tau_0 = 10^{-4}$.
  (Inset) Magnitudes of the Fourier coefficients $a^{(n)}$ and $b^{(n)}$ against $\gamma_0$ for $\phi=0.87$ and $\omega \tau_0 = 10^{-4}$ with $n=1$.
  The dashed line represents $a^{(n)}, b^{(n)} \propto \gamma_0$.
}
\label{Fourier_GR}
\end{figure}

Figure \ref{Fourier_GR} plots the magnitude of the Fourier coefficients $a^{(n)}$ and $b^{(n)}$ against $n$ for $\phi=0.87$ and $\gamma_0 = 0.02$ with $\omega \tau_0 = 10^{-4}$.
As in the overdamped system, $a^{(1)}$ takes the largest value, and the other components are non-zero.
In the inset of Fig. \ref{Fourier_GR}, we show $a^{(1)}$ and $b^{(1)}$ against $\gamma_0$ for $\phi=0.87$ and $\gamma_0 = 0.02$ with $\omega \tau_0 = 10^{-4}$, which are proportional to $\gamma_0$.

\begin{figure}[htbp]
\includegraphics[width=0.7\linewidth]{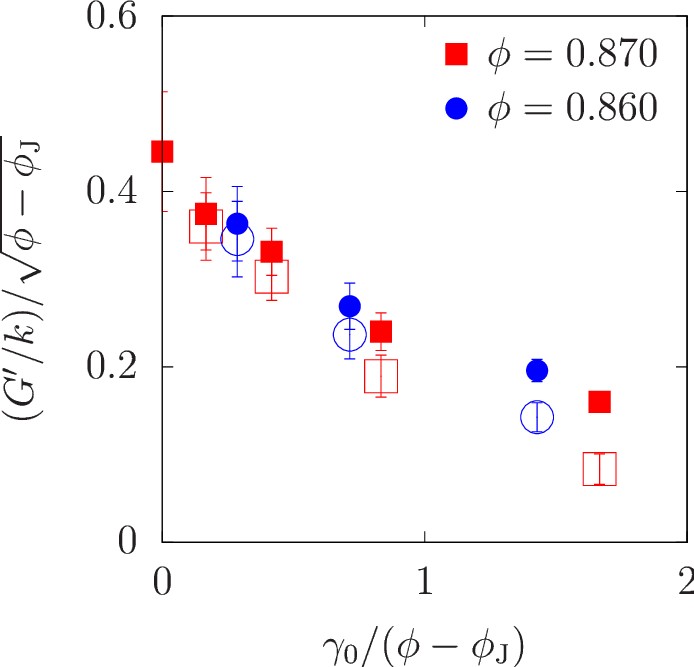}
  \caption{Scaled storage modulus $ G'/\sqrt{\phi - \phi_J}$ of underdamped particles (filled symbols) and its theoretical expression using $G'_{\rm T}$ (open symbols) in Eq. (14) against scaled $\gamma_0/(\phi - \phi_J)$ in the absorbing state scaled by the distance $\phi - \phi_J$ from the jamming point for $\omega = 10^{-4}\tau_0^{-1}$ with $\phi=0.870$ and $0.860$.
}
\label{Gp_GR}
\end{figure}

In Fig. \ref{Gp_GR}, we plot the scaled storage modulus $G'$ of the underdamped particles in the absorbing state against the scaled amplitude $\gamma_0$ for $\omega = 10^{-4}\tau_0^{-1}$ with $\phi=0.870$ and $0.860$.
The storage modulus exhibits SAS.
The corresponding theoretical expression $G'_{\rm T}$ in Eq. (14) as open symbols is also presented in Fig. \ref{Gp_GR}, which quantitatively reproduces the numerical results except for the region of quite large $\gamma_0$

\begin{figure}[htbp]
\includegraphics[width=0.9\linewidth]{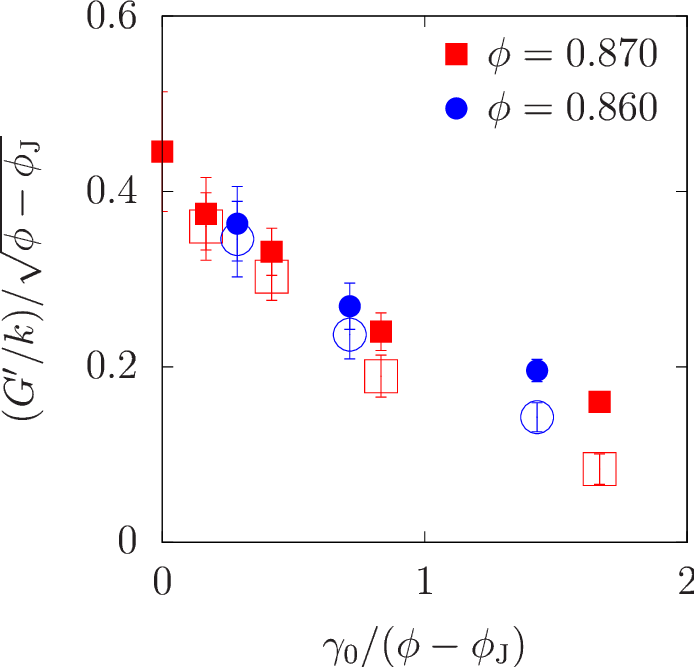}
  \caption{(a) Loss modulus $G''$ of underdamped particles obtained in our simulation (filled symbols) and its theoretical expression $G''_{\rm T}$ (open symbols) in Eq. (14) against $\gamma_0$ for $\omega = 10^{-4}\tau_0^{-1}$ with $\phi=0.870$ and $0.860$.
  (b) Loss modulus $G''$ against $\omega$ for $\phi=0.87$ with $\gamma_0=0.01$.
}
\label{Gpp_GR}
\end{figure}

In Fig. \ref{Gpp_GR}(a), we present the loss modulus $G''$ obtained in our simulation and its theoretical expression $G''_{\rm T}$ in Eq. (15) against $\gamma_0$ for $\omega = 10^{-4}\tau_0^{-1}$ with $\phi=0.870$ and $0.860$ in the underdamped system.
The loss modulus $G''$ does not strongly depend on $\phi$ and $\gamma_0$, and the theoretical expression $G''_{\rm T}$ reproduces the numerical results.
In Fig. \ref{Gpp_GR}(b), we plot the loss modulus $G''$ against $\omega$ for $\phi=0.87$ with $\gamma=0.01$.
The loss modulus seems to converge to a non-zero value in the limit $\omega \to 0$.

The results in this section are consistent with those in the main text for the overdamped system.
This indicates that the results presented in the main text are universal for jammed disordered materials.

\section{Dependence of $G'$ and $G''$ on $N$ and $n_c$}
\label{Sec:Nc}

\begin{figure}[htbp]
  \begin{center}
    \begin{tabular}{c}
      \begin{minipage}{0.5\hsize}
      \begin{center}
        \includegraphics[width=1.0\linewidth]{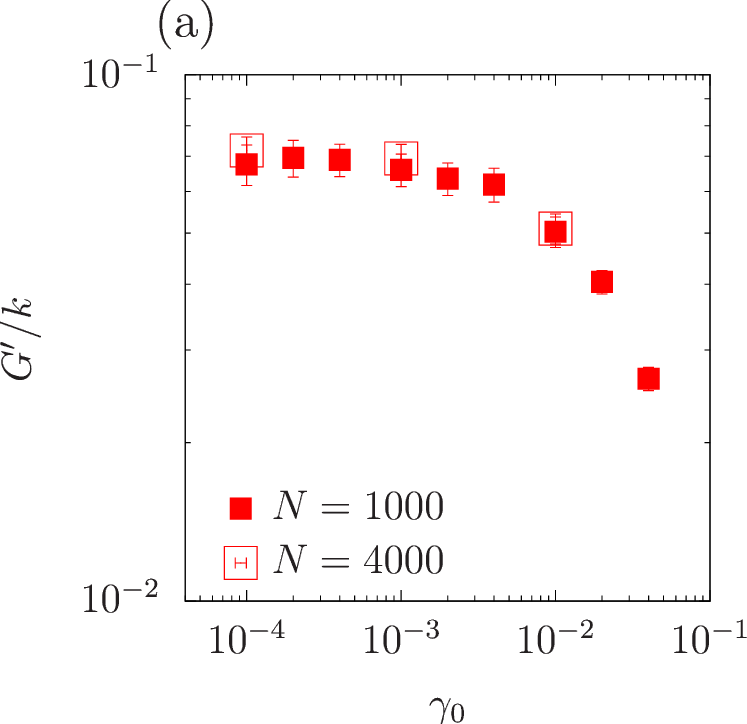}
      \end{center}
      \end{minipage}
      \begin{minipage}{0.5\hsize}
      \begin{center}
        \includegraphics[width=0.9\linewidth]{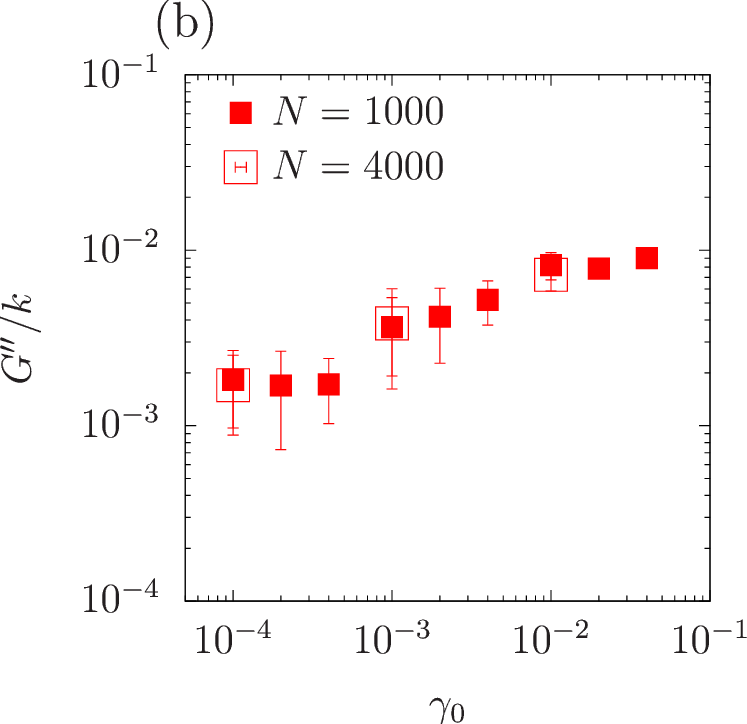}
      \end{center}
      \end{minipage}
    \end{tabular}
    \caption{(a) Storage modulus $G'$ against $\gamma_0$ for $\omega = 10^{-4} \tau_0^{-1}$, $\phi = 0.870$, and $n_c = 20$ with $N=1000$ and $4000$.
  (b) Loss modulus $G''$ against $\gamma_0$ for $\omega = 10^{-4} \tau_0^{-1}$, $\phi = 0.870$, and $n_c = 20$ with $N=1000$ and $4000$.
}
\label{G_N4000}
  \end{center}
\end{figure}

In this section, we show the dependence of $G'$ and $G''$ on the numbers of particles $N$ and cycles $n_c$ for the overdamped dynamics discussed in the main text.
In Figs. \ref{G_N4000}(a) and (b), we plot $G'$ and $G''$ against $\gamma_0$ for $\omega = 10^{-4} \tau_0^{-1}$, $\phi=0.870$, and $n_c = 20$ with $N=1000$ and $4000$, respectively.
The shear moduli $G'$ and $G''$ for $N=1000$ and $4000$ are consistent within error bars.

\begin{figure}[htbp]
  \begin{center}
    \begin{tabular}{c}
      \begin{minipage}{0.5\hsize}
      \begin{center}
        \includegraphics[width=1.0\linewidth]{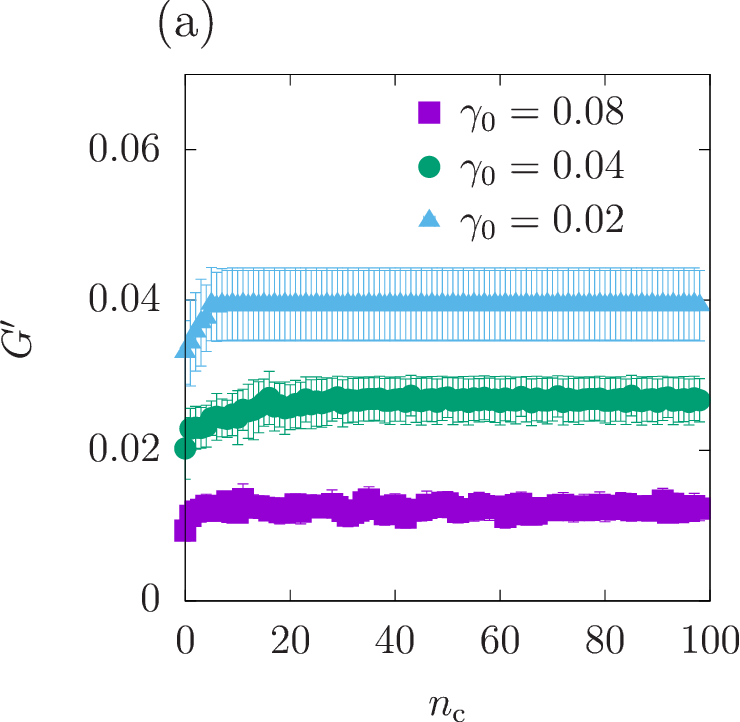}
      \end{center}
      \end{minipage}
      \begin{minipage}{0.5\hsize}
      \begin{center}
        \includegraphics[width=0.9\linewidth]{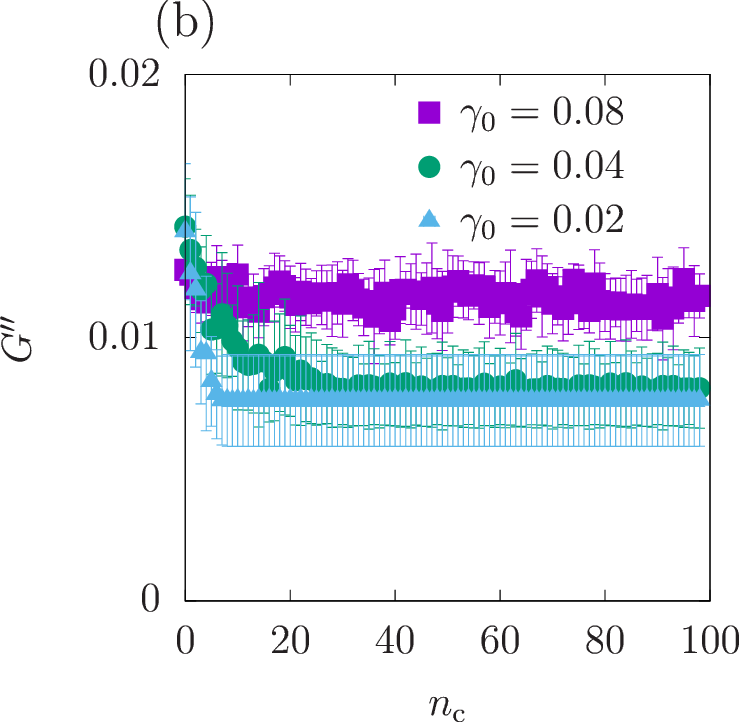}
      \end{center}
      \end{minipage}
    \end{tabular}
    \caption{(a) Storage modulus $G'$ against $n_c$ for $\omega = 10^{-4} \tau_0^{-1}$, $\phi = 0.870$, and $N=1000$ with $\gamma_0 = 0.02, 0.04$, and $0.08$.
  (b) Loss modulus $G''$ against $n_c$ for $\omega = 10^{-4} \tau_0^{-1}$, $\phi = 0.870$, and $N=1000$ with $\gamma_0 = 0.02, 0.04$, and $0.08$.
}
\label{G_n}
  \end{center}
\end{figure}

Figures \ref{G_n}(a) and (b) show $G'$ and $G''$ against $n_c$ for $\omega = 10^{-4} \tau_0^{-1}$, $\phi=0.870$, and $N=1000$ with $\gamma_0 = 0.02, 0.04$, and $0.08$, respectively.
The shear moduli $G'$ and $G''$ reach statistical steady states for $n_c \ge 20$ within error bars.

\section{Particle displacement and yielding strain amplitude}
\label{Sec:yield}

\begin{figure}[htbp]
\includegraphics[width=1.0\linewidth]{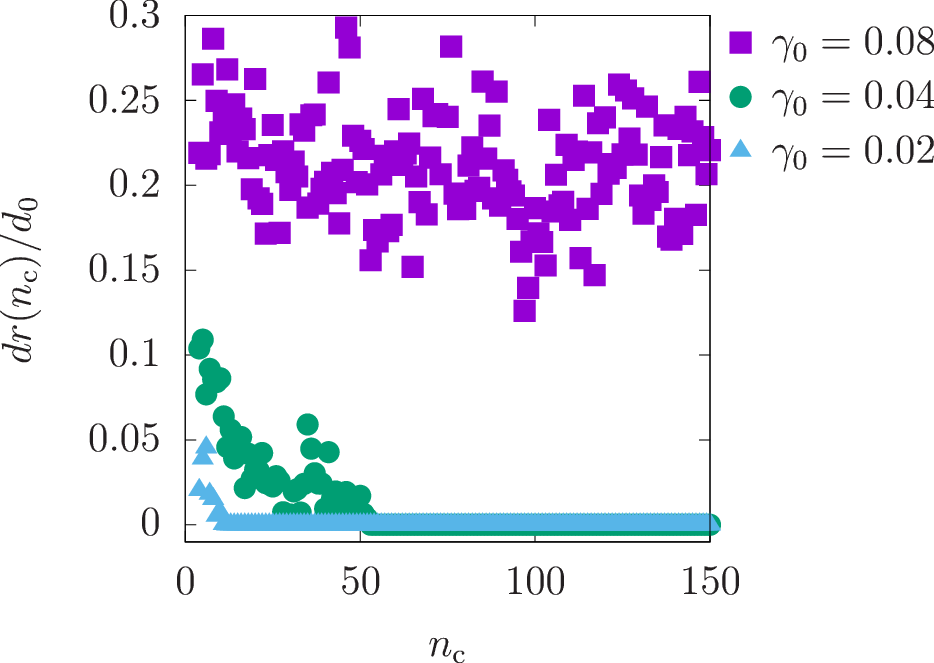}
  \caption{ 
  Displacements of particles $dr(n_c)$ against $n_c$ for $\omega = 10^{-4} \tau_0^{-1}$ and $\phi=0.860$ with $\gamma_0 = 0.08, 0.04$, and $0.02$.
}
\label{drn_OD_n}
\end{figure}

In this section, we show the time evolution of the displacements of particles before reaching the absorbing state and the evaluation of the yielding strain amplitude $\gamma_c$.
Here, we introduce the particle displacement between $n_c$-th and $(n_c-m)$-th cycles as
\begin{equation}
  dr_m(n_c) = \sum_{i=1}^N \left |\boldsymbol r_i(n_cT) - \boldsymbol r_i((n_c-m)T) \right | /N
\end{equation}
with the period $T = 2\pi/\omega$.
We define $dr(n_c)$ as
the minimum value of $dr_m(n_c)$ for $m$.
In Fig. \ref{drn_OD_n}, we plot $dr(n_c)$ against $n_c$ for $\omega = 10^{-4} \tau_0^{-1}$ and $\phi=0.860$ with $\gamma_0 = 0.08, 0.04$, and $0.02$.
For $\gamma_0 = 0.08$, $dr(n_c)$ remains non-zero, while it approaches $0$ after a transient for $\gamma_0 = 0.04$ and $0.02$.
It is noteworthy that $dr(n_c)$ reaches a steady state for $n_c>50$ in the case of $\gamma_0=0.04$, which is much larger than the steady $n_c$ in Fig. S6.

\begin{figure}[htbp]
\includegraphics[width=0.8\linewidth]{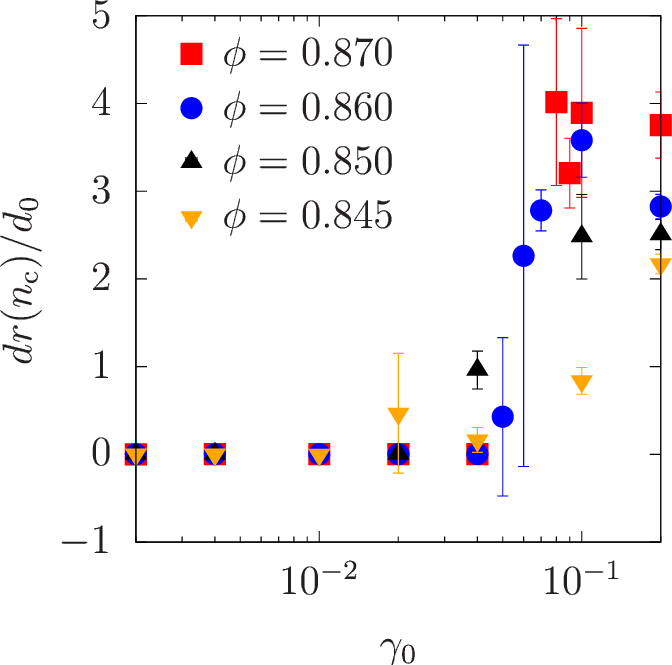}
  \caption{
    Displacement of particles $dr(n_c)$ against $\gamma_0$ at $n_c=100$ for $\omega = 10^{-4} \tau_0^{-1}$ with $\phi = 0.870, 0.860, 0.850$ and $0.845$.
}
\label{drn_OD}
\end{figure}

In Fig. \ref{drn_OD}, we plot $dr(n_c)$ against $\gamma_0$ at $n_c=100$ for $\omega = 10^{-4} \tau_0^{-1}$ with $\phi = 0.870, 0.860, 0.850$ and $0.845$.
For all $\phi$, $dr(n_c)$ changes from $0$ to non-zero values as $\gamma_0$ increases.
We call the absorbing state for $dr(n_c)<d_c$ with smaller $\gamma_0$ and the plastic state for $dr(n_c)>d_c$ with larger $\gamma_0$.
The yielding strain amplitude $\gamma_c$ is defined as the boundary between these states.
From Fig. \ref{drn_OD}, we estimate $0.04 < \gamma_c < 0.08$ for $\phi=0.870$, $0.04 < \gamma_c < 0.05$ for $\phi=0.860$, $0.02 < \gamma_c < 0.03$ for $\phi=0.850$, and $0.01 < \gamma_c < 0.02$ for $\phi=0.845$.

\section{Stress-strain curve}
\label{Sec:stress}

\begin{figure}[htbp]
\includegraphics[width=0.9\linewidth]{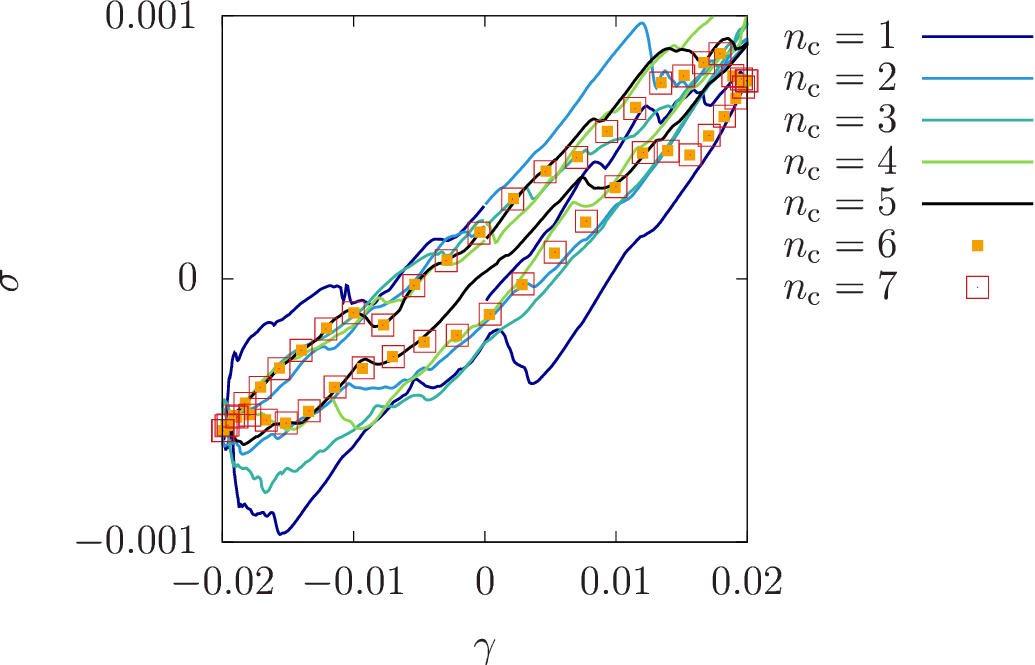}
  \caption{
Plots of shear stress $\sigma$ against $\gamma$ for $\gamma_0 = 0.02$, $\omega=10^{-4}\tau_0^{-1}$, and $\phi=0.87$ corresponding to $\phi - \phi_J=0.029$ with various $n_c$.
}
\label{s_ga_re}
\end{figure}

\begin{figure}[htbp]
\includegraphics[width=0.9\linewidth]{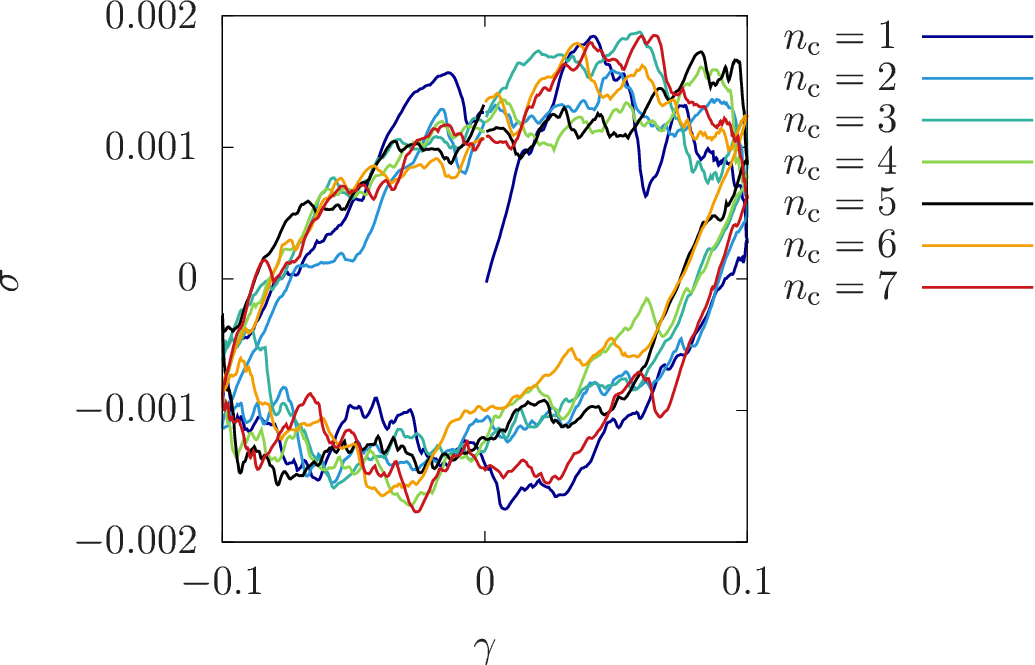}
  \caption{
Plots of shear stress $\sigma$ against $\gamma$ for $\gamma_0 = 0.1$, $\omega=10^{-4}\tau_0^{-1}$ and $\phi=0.87$ corresponding to $\phi - \phi_J=0.029$ with various $n_c$.
}
\label{s_ga_ir}
\end{figure}

In this section, we present typical stress strain curves in the absorbing and plastic states including their time evolution.
Figure \ref{s_ga_re} displays the shear stress $\sigma$ against the strain $\gamma$ with $\gamma_0 = 0.02$ for different $n_c$.
For $n_c \le 5$, the stress-strain curves are not convergent, which indicate the system is in a transient state.
For $n_c = 6$ and $7$, the stress-strain curves become identical in the absorbing state.
We plot the shear stress $\sigma$ against the strain $\gamma$ for $\gamma_0 = 0.1$ in Fig. \ref{s_ga_ir}.
All the stress-strain curves are different for all $n_c$ because the system is in the plastic state.

\section{Dependence of trajectories on $\gamma_0$ and $\omega$}
\label{Sec:LoopDep}

\begin{figure}[htbp]
  \begin{center}
    \begin{tabular}{c}
      \begin{minipage}{0.5\hsize}
      \begin{center}
        \includegraphics[width=1.0\linewidth]{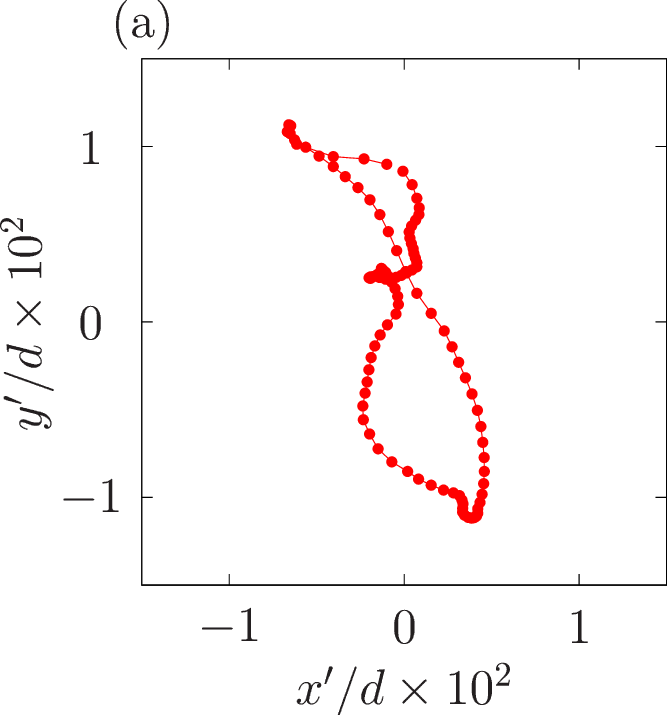}
      \end{center}
      \end{minipage}
      \begin{minipage}{0.50\hsize}
      \begin{center}
        \includegraphics[width=1.0\linewidth]{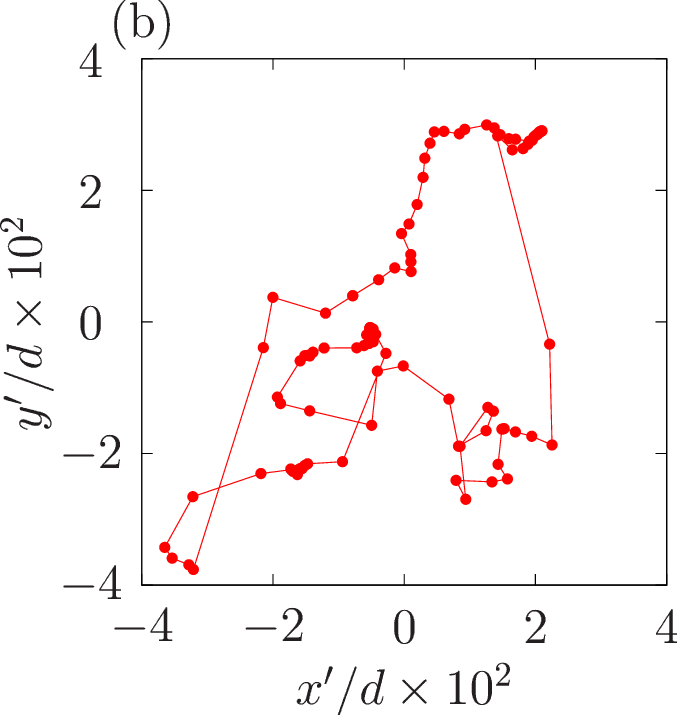}
      \end{center}
      \end{minipage}
    \end{tabular}
\caption{
  Non-affine particle trajectories in the last cycle for $\omega = 10^{-3}\tau_0^{-1}$ (a) and $10^{-5}\tau_0^{-1}$ (b) with $\phi=0.87$ and $\gamma_0=0.01$.
}
\label{trjna_ga0.01}
  \end{center}
\end{figure}

In this section, we present how particle trajectories depend on $\gamma_0$ and $\omega$.
In Fig. \ref{trjna_ga0.01}, we plot the non-affine particle trajectories in the last cycle
for $\omega = 10^{-3}\tau_0^{-1}$ and $10^{-5}\tau_0^{-1}$ with $\phi=0.87$ and $\gamma_0=0.01$.
Let us introduce
\begin{equation}
  {\boldsymbol r}'_i = (x'_i, y'_i) =
  \tilde {\boldsymbol r}_i - {\boldsymbol R}_i.
\end{equation}
The trajectories form nontrivial loops, which remain for smaller $\omega$.

\begin{figure}[htbp]
  \begin{center}
    \begin{tabular}{c}
      \begin{minipage}{0.5\hsize}
      \begin{center}
        \includegraphics[width=1.0\linewidth]{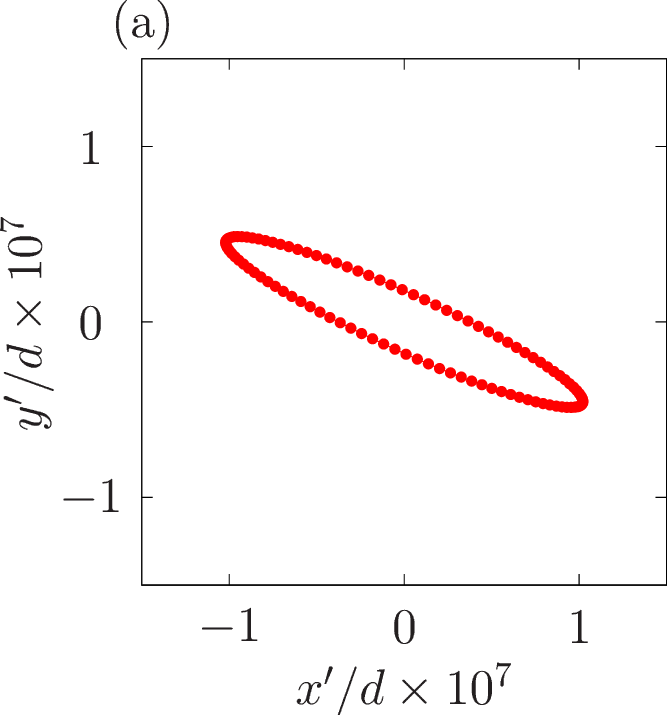}
      \end{center}
      \end{minipage}
      \begin{minipage}{0.50\hsize}
      \begin{center}
        \includegraphics[width=1.0\linewidth]{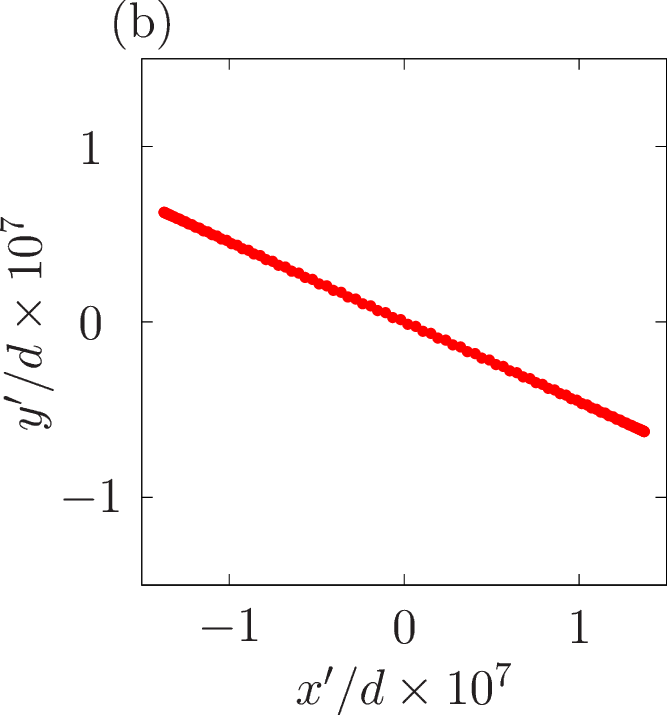}
      \end{center}
      \end{minipage}
    \end{tabular}
\caption{
  Non-affine particle trajectories in the last cycle for $\omega = 10^{-3}\tau_0^{-1}$ (a) and $10^{-5}\tau_0^{-1}$ (b) with $\phi=0.87$ and $\gamma_0=1.0 \times 10^{-7}$.
}
\label{trjna_ga0.0000001}
  \end{center}
\end{figure}

Figure \ref{trjna_ga0.0000001} represents the non-affine particle trajectories in the last cycle
for $\omega = 10^{-3}\tau_0^{-1}$ and $10^{-5}\tau_0^{-1}$ with $\phi=0.87$ and $\gamma_0=1.0 \times 10^{-7}$.
In Fig. \ref{trjna_ga0.0000001} (a), the trajectory with $\omega = 10^{-3} \tau_0^{-1}$ form an ellipse, but the trajectory becomes a straight line for $\omega = 10^{-5} \tau_0^{-1}$ in Fig \ref{trjna_ga0.0000001} (b).

\section{Effect of trajectories with longer periods}
\label{Sec:double}

In this section, we discuss the effect of closed trajectories with periods longer than $2\pi$.
As indicated by Refs. [27, 28, 30, 31, 33], some samples exhibit non-trivial absorbing trajectories where particles return to their original positions after more than one cycle of oscillatory shear.
In these samples, the non-affine trajectories of a particle $\boldsymbol r_i(\theta)$ satisfy
\begin{equation}
\boldsymbol r_i(\theta) = \boldsymbol r_i(\theta + 2 M \pi)
\end{equation}
with $M= 2,3,4, \cdots$.
In this case, $\boldsymbol r_i(\theta)$ for $0 < \theta < 2M \pi$ is expressed in the Fourier series as
\begin{equation}
  \tilde {\boldsymbol r}_i(\theta) =   \boldsymbol R'_i +
  \sum_{m=1}^\infty \left ( \boldsymbol A_i^{(m)} \sin \frac{m \theta}{M}
  + \boldsymbol B_i^{(m)} \cos \frac{m \theta}{M} \right )
  \label{eq:Fourier_M}
\end{equation}
with
\begin{eqnarray}
  \label{eq:Rd}
  \boldsymbol R'_i  & = & \frac{1}{2M\pi}
  \int_0^{2M\pi} \ d\theta \ \tilde{\boldsymbol r}_i(\theta),
\end{eqnarray}
and the Fourier coefficients
\begin{eqnarray}
  \label{eq:A}
  \boldsymbol A_i^{(m)}  & = & \frac{1}{M\pi}
  \int_0^{2M\pi} \ d\theta \ \sin \frac{m \theta}{M} \ \tilde{\boldsymbol r}_i(\theta), \\
  \label{eq:B}
  \boldsymbol B_i^{(m)}  & = & \frac{1}{M\pi}
  \int_0^{2M\pi} \ d\theta \ \cos \frac{m\theta}{M} \ \tilde{\boldsymbol r}_i(\theta).
\end{eqnarray}

However, in Eqs. (4) and (5), we need $\boldsymbol r_i(\theta)$ for $0 \le \theta < 2 \pi$ to calculate $G'$ and $G''$.
When $\boldsymbol r_i(\theta)$ is restricted to $0 \le \theta < 2 \pi$, we can use Eq. (8) with the Fourier coefficient given by Eqs. (10) and (11) as an expression of the trajectory, and we obtain the theoretical expressions Eqs. (14) and (15) even in this case.
It should be noted that samples in the absorbing state with longer periods are rare, and the probability of emerging such a trajectory is smaller than $0.01$ for sufficiently packed systems above the jamming point, as shown in Ref. [30].
Therefore, we can ignore the effect of rare samples.

\section{Loss modulus}
\label{Sec:Loss}

\begin{figure}[htbp]
\includegraphics[width=0.9\linewidth]{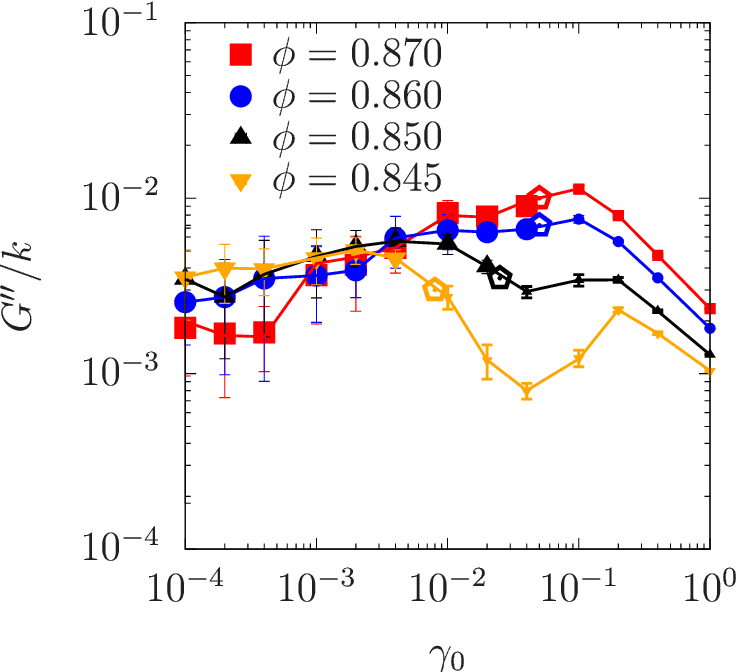}
  \caption{ Loss modulus $G''$ against $\gamma_0$ for $\omega = 10^{-4}\tau_0^{-1}$ with $\phi=0.870$ and $0.860$.
  The larger (smaller) filled symbols represent the data in the absorbing (plastic) state.
The open pentagons represent the yield strain amplitude $\gamma_c$.
}
\label{Gpp_full}
\end{figure}

In Fig. \ref{Gpp_full}, we plot the loss modulus $G''$ against $\gamma_0$ for $\omega = 10^{-4}\tau_0^{-1}$ with $\phi=0.870$ and $0.860$ including the data in the absorbing and plastic states.
This figure corresponds to Fig. 3(a) in the main text, but Fig. S13 contains the data for a wide range of $\gamma_0$.
The previous studies [21, 22] reported that the loss modulus has a peak around the yield strain for an underdamped system, but the peak of $G''$ is not clearly visible in our overdamped system.

\section{Shear modulus for small $\gamma_0$} 
\label{Sec:KV}

\begin{figure}[htbp]
\includegraphics[width=0.7\linewidth]{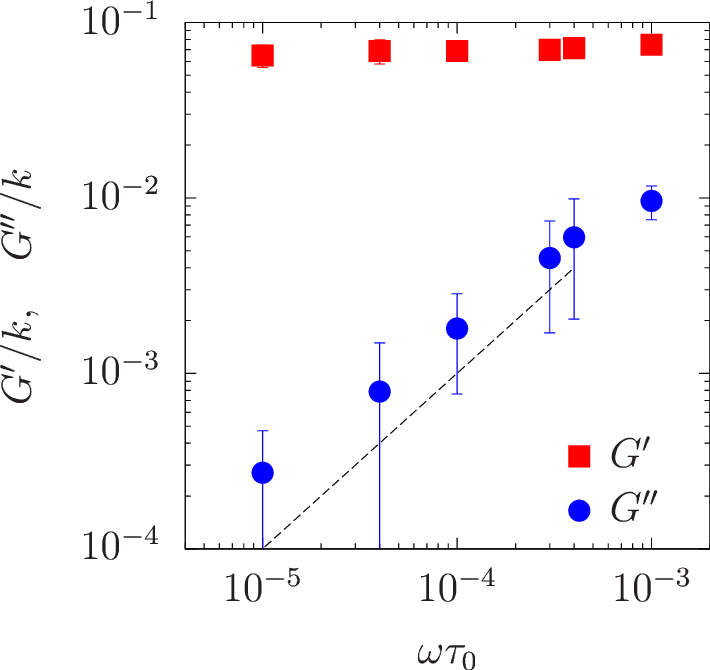}
  \caption{Plots of $G'$ and $G''$ against $\omega$ for $\phi=0.870$ and $\gamma_0 = 1.0 \times 10^{-7}$.
  The dashed line represents $G'' \propto \omega$.
}
\label{G_KV}
\end{figure}

In this section, we demonstrate that $G'$ and $G''$ obey the Kelvin--Voigt model for a sufficiently small $\gamma_0$.
Figure \ref{G_KV} is a set of plots of $G'$ and $G''$ against $\omega \tau_0$ for $\phi=0.870$ and $\gamma_0 = 1.0 \times 10^{-7}$, where $G'$ is almost independent of $\omega$ and $G''$ is proportional to $\omega$.
This behavior is consistent with that of the Kelvin--Voigt model.

\section{Relationship between closed trajectories and the Fourier coefficients}
\label{loop}

\begin{figure}[htbp]
\includegraphics[width=1.0\linewidth]{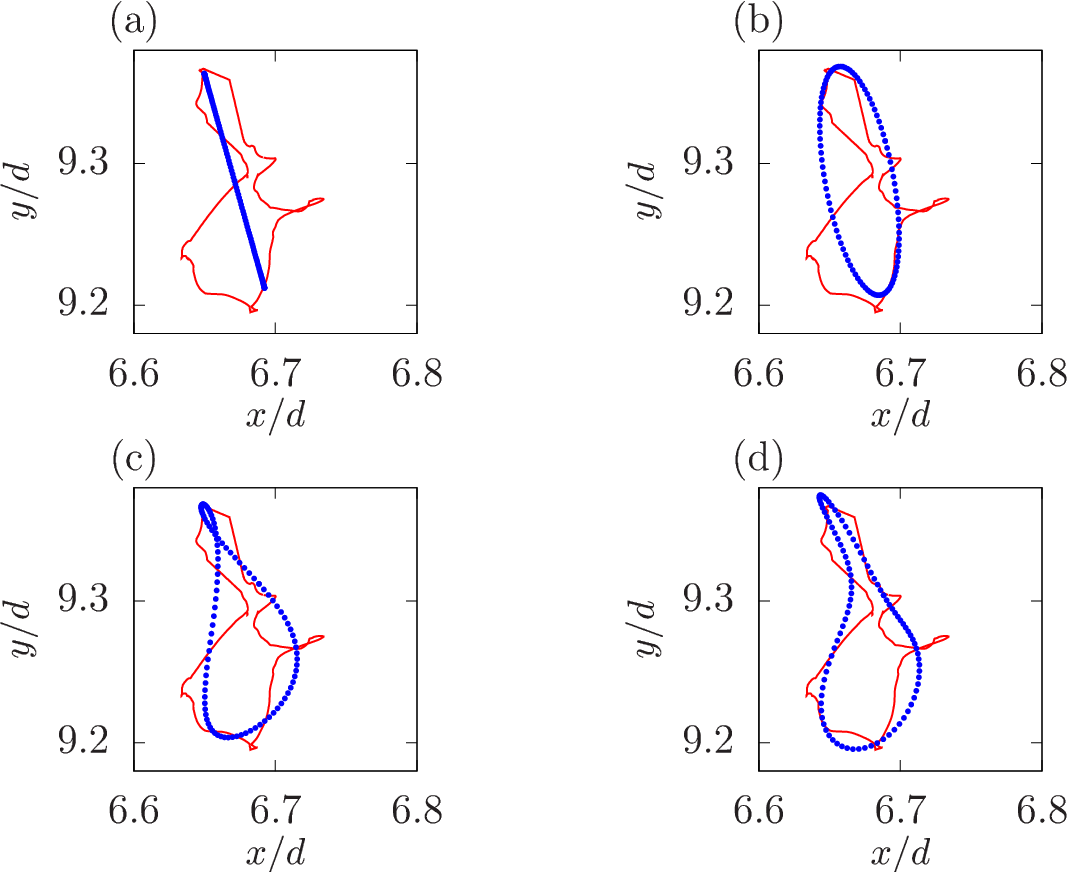}
  \caption{Trajectory shown in Fig. 1(a) of the main text and its approximate trajectories with some restricted modes. The red solid lines represent the original data, and the blue filled circles represent the approximate trajectory using (a) $\boldsymbol a_i^{\rm (1)}$, (b) $\boldsymbol a_i^{\rm (1)}$ and $\boldsymbol b_i^{\rm (1)}$, (c) $\boldsymbol a_i^{\rm (n)}$ and $\boldsymbol b_i^{\rm (n)}$ with $n=1$ and $2$, and (d) $\boldsymbol a_i^{\rm (n)}$ and $\boldsymbol b_i^{\rm (n)}$ with $n=1,2,$ and $3$.
}
\label{trjna_F}
\end{figure}

In this section, we present how the trajectory of a particle depends on the Fourier coefficients.
Figure \ref{trjna_F} compares the trajectory of a particle corresponding to Fig. 1(a) 
with its approximate trajectory using Eq. (8) with some restricted modes, where we estimate the coefficients using the true trajectory.
In Fig. \ref{trjna_F} (a), we plot the approximate trajectory (blue filled circles) using only $\boldsymbol a_i^{\rm (1)}$, where we set the other coefficients to $0$.
The approximate trajectory (blue filled circles) is a straight line.
Figure \ref{trjna_F} (b) shows the approximate trajectory using $\boldsymbol a_i^{\rm (1)}$ and $\boldsymbol b_i^{\rm (1)}$, where the trajectory becomes an ellipse.
As we increase the number of modes, the approximate trajectory approaches the true trajectory, as shown in Figs. \ref{trjna_F} (c) and (d).

\section{Cyclic contact changes}
\label{Sec:CC}

\begin{figure}[htbp]
\includegraphics[width=1.0\linewidth]{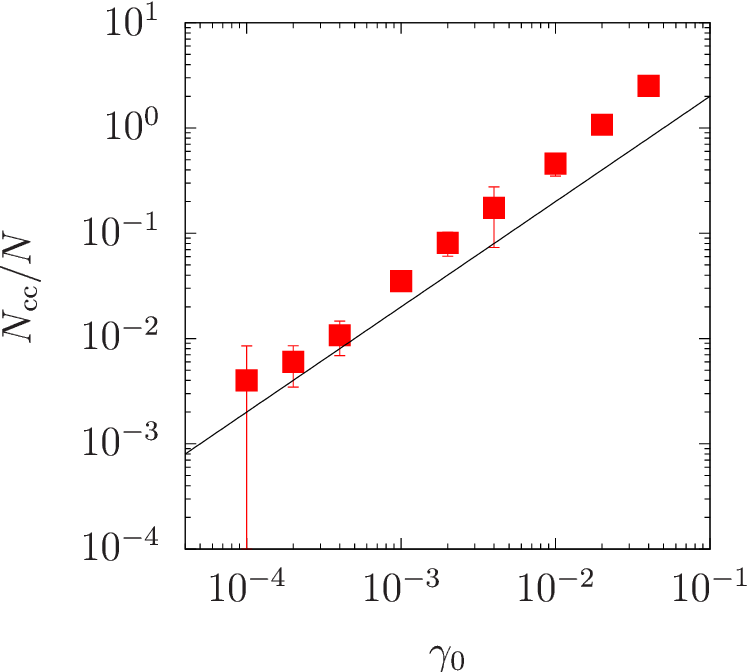}
  \caption{
    The number of cyclic contact changes $N_{\rm cc}$ during the last cycle for $\omega = 10^{-4} \tau_0^{-1}$ with $\phi=0.870$ against $\gamma_0$.
    The solid line represents $N_{\rm cc} \sim \gamma_0$.
}
\label{OC}
\end{figure}

In this section, we show the number of contact changes during the last cycle in the absorbing state.
References [21, 23, 25, 26] demonstrate that the nontrivial loops originate from cyclic open and close contacts.
Here, we define $N_{cc}$ as the number of events where the same contact opens and closes again during the last cycle.
In Fig. \ref{OC}, we present $N_{\rm cc}$ during the last cycle for $\omega = 10^{-4} \tau_0^{-1}$ with $\phi=0.870$ against $\gamma_0$ in the absorbing state.
The number of cyclic contact changes $N_{\rm cc}$ is nearly proportional to $\gamma_0$.
This dependence is consistent with the behaviors of $a^{(n)}$ and $b^{(n)}$ of the Fourier components, which are almost proportional to $\gamma_0$.

\section{Details of the theoretical analysis}
\label{Sec:theory}

In this section, we derive Eqs. (14) and (15) in the main text by assuming
$|a_i^{(n)}| \sim |b_i^{(n)}| \sim \gamma_0$ and $\gamma_0 \ll 1$.
From Eq. (13) in the main text, $x_{ij}(\theta)$ and $y_{ij}(\theta)$ are given by
\begin{align}
  \label{xij}
  &  x_{ij}  =  X_{ij} + \gamma_0 \sin \theta Y_{ij}
   +\sum_{n=1}^\infty \left ( a_{ij,x}^{(n)} \sin n \theta
  + b_{ij,x}^{(n)} \cos n \theta \right ), \\
  & y_{ij}  =  Y_{ij}
  +\sum_{n=1}^\infty \left ( a_{ij,y}^{(n)} \sin n \theta
  + b_{ij,y}^{(n)} \cos n \theta \right ),
\end{align}
where
$\boldsymbol a^{(n)}_{ij} =  (a_{ij,x}^{(n)},a_{ij,y}^{(n)}) = \boldsymbol a^{(n)}_{i}-\boldsymbol a^{(n)}_{j}$, $\boldsymbol b^{(n)}_{ij}  =  (b_{ij,x}^{(n)},b_{ij,y}^{(n)})=  \boldsymbol b^{(n)}_{i}-\boldsymbol b^{(n)}_{j}$.
Using this equation and neglecting the terms of $O(\gamma_0^2)$,
$|\boldsymbol r_{ij}(\theta)|^2 = x_{ij}^2 +  y_{ij}^2 $ is given by
\begin{eqnarray}
 |\boldsymbol r_{ij}(\theta)|^2 \simeq
   R_{ij}^2 \left \{ 1 + 2E_{ij}(\theta)\right \}
   \label{r2}
\end{eqnarray}
with
\begin{eqnarray}
  E_{ij}(\theta)  & = &  
  \sum_{n=1}^\infty \frac{\boldsymbol R_{ij} \cdot \boldsymbol a_{ij}^{(n)}}{R_{ij}^2} \sin n \theta
  + \sum_{n=1}^\infty \frac{\boldsymbol R_{ij}\cdot  \boldsymbol b_{ij}^{(n)}}{R_{ij}^2} \cos n \theta \nonumber \\
&  &+
\gamma_0 \frac{X_{ij}Y_{ij}}{R_{ij}^2} \sin \theta.
  \label{eq:E}
\end{eqnarray}
From Eq. \eqref{r2}, $r_{ij}(\theta)$ is approximately obtained as
\begin{eqnarray}
  r_{ij}(\theta) \simeq
   R_{ij} \left \{ 1 + E_{ij}(\theta)\right \}
   \label{ex:r}
\end{eqnarray}
up to $O(\gamma_0)$.
Using this equation, we obtain $\Psi(r) = -U'(r)/r$ up to $O(\gamma_0)$ as
\begin{eqnarray}
  \Psi(r_{ij}(\theta)) & \simeq &
  \Psi(R_{ij}) + \Psi'(R_{ij}) R_{ij}E_{ij}(\theta).
  \label{ex:Psi}
\end{eqnarray}

Substituting Eqs. \eqref{xij}--\eqref{ex:Psi} into Eq. (6), we obtain
\begin{eqnarray}
  \sigma(\theta)  & = &
  - \frac{1}{L^2}  \sum_{(i,j)}
  \left \{ \Psi(R_{ij}) + \Psi'(R_{ij}) R_{ij}E_{ij}(\theta) \right \}
  \nonumber \\
  & & \times 
\left \{ X_{ij} + \gamma_0 \sin \theta Y_{ij}
   +\sum_{n=1}^\infty \left ( a_{ij,x}^{(n)} \sin n \theta
  + b_{ij,x}^{(n)} \cos n \theta \right ) \right \} \nonumber \\ 
  & & \times \left \{  Y_{ij}
  +\sum_{n=1}^\infty \left ( a_{ij,y}^{(n)} \sin n \theta
  + b_{ij,y}^{(n)} \cos n \theta \right ) \right \}.
\end{eqnarray}
Here, we abbreviate $\displaystyle \sum_i \sum_{j>i}$ as $\displaystyle \sum_{(i,j)}$.
Neglecting the terms of $O(\gamma_0^2)$, $\sigma(\theta)$ is approximated as
\begin{eqnarray}
  \sigma(\theta) 
  &\simeq & - \frac{1}{L^2} \sum_{(i,j)} X_{ij}Y_{ij}\Psi(R_{ij})
 - \frac{1}{L^2}\sum_{(i,j)}
  \gamma_0 \sin \theta Y_{ij}^2 \Psi(R_{ij}) \nonumber \\
  & & - \frac{1}{L^2}\sum_{(i,j)}\sum_{n=1}^\infty \left ( a_{ij,x}^{(n)} \sin n \theta 
  + b_{ij,x}^{(n)} \cos n \theta \right ) Y_{ij} \Psi(R_{ij}) \nonumber \\
  & & - \frac{1}{L^2}\sum_{(i,j)}\sum_{n=1}^\infty X_{ij} \left ( a_{ij,y}^{(n)} \sin n \theta 
  + b_{ij,y}^{(n)} \cos n \theta \right ) \Psi(R_{ij}) \nonumber \\
  & & - \frac{1}{L^2}\sum_{(i,j)}X_{ij}Y_{ij}\Psi'(R_{ij}) R_{ij}E_{ij}(\theta).
\end{eqnarray}
By substituting this equation into Eqs. (4) and (5) and using
\begin{align}
  & \frac{1}{\pi} \int_0^{2\pi} \ d\theta \  \sin m\theta \sin n \theta  = \delta_{mn}, \\
  & \frac{1}{\pi} \int_0^{2\pi} \ d\theta \  \sin m\theta \cos n \theta  = 0,
\end{align}
we obtain Eqs. (14) and (15) in the main text.

\section{Components of shear moduli}
\label{Sec:Components}

\begin{figure}[htbp]
\includegraphics[width=0.9\linewidth]{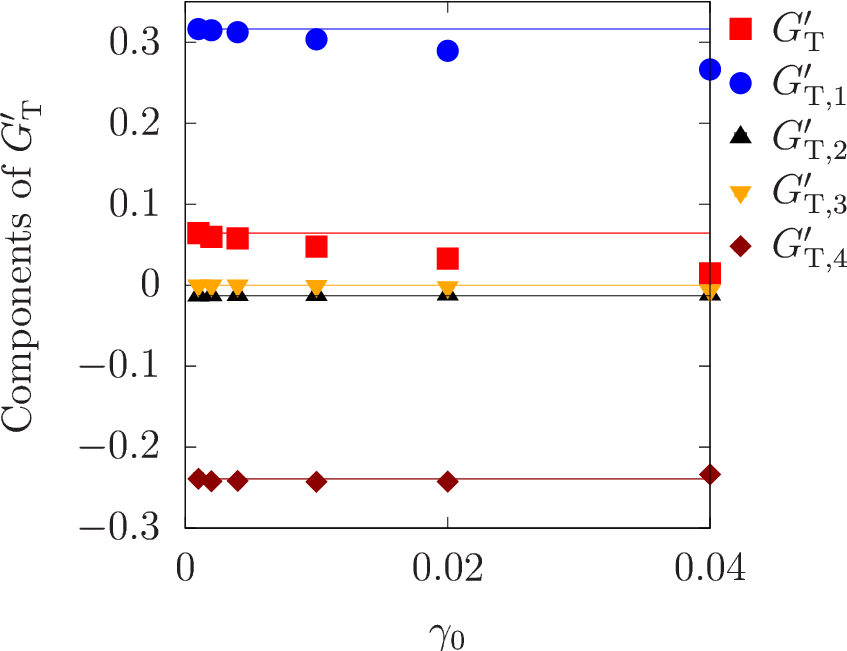}
  \caption{ $G'_{{\rm T},n}$ with $n=1,2,3$ and $4$ against $\gamma_0$ in the absorbing state for $\omega = 10^{-4}\tau_0^{-1}$ with $\phi=0.870$.
The horizontal lines represent $G'_{{\rm T},n}$ in the limit $\gamma_0 \to 0$, which is estimated at $\gamma_0 = 0.001$.
}
\label{Gp_components}
\end{figure}

In this section, we clarify what terms of the theoretical expressions $G'_{\rm T}$ and $G''_{\rm T}$ in the absorbing state in Eqs. (14) and (15) are dominant.
Here, $G'_{\rm T}$ consists of four terms as
\begin{eqnarray}
  G'_{\rm T} & = & G'_{{\rm T},1} +  G'_{{\rm T},2} + G'_{{\rm T},3} +  G'_{{\rm T},4} 
\end{eqnarray}
with
\begin{align}
  G'_{{\rm T},1} & =  -  \frac{1}{L^2}\sum_{i,j} \left \langle
\frac{X_{ij}^2 Y_{ij}^2 }{R_{ij}}\Psi'(R_{ij}) 
   \right \rangle, \\
  G'_{{\rm T},2} &=  
  - \frac{1}{L^2}\sum_{i,j} \left \langle 
Y_{ij}^2 \Psi(R_{ij})  
  \right \rangle, \\
   G'_{{\rm T},3}  &= - \frac{1}{L^2}\sum_{i,j} \left \langle  \left ( \frac{a_{ij,x}^{(1)}}{\gamma_0}Y_{ij}  + X_{ij} \frac{a_{ij,y}^{(1)}}{\gamma_0} \right ) \Psi(R_{ij})  \right \rangle, \\
    G'_{{\rm T},4} &= - \left \langle \frac{1}{L^2}\sum_{i,j}X_{ij}Y_{ij}\Psi'(R_{ij})
   \frac{\boldsymbol R_{ij} \cdot \boldsymbol a_{ij}^{(1)}}{\gamma_0 R_{ij}} \right \rangle,
\end{align}
where $G'_{{\rm T},1}$ and $G'_{{\rm T},2}$ represent the contributions from the affine motion, respectively, while $G'_{{\rm T},1}$ and $G'_{{\rm T},2}$ are the contributions from the non-affine motion, respectively.
In Fig. \ref{Gp_components}, we show $G'_{{\rm T},n}$ in the absorbing state for $\omega = 10^{-4}\tau_0^{-1}$ with $\phi=0.870$.
We find that $G'_{{\rm T},1}$ and $G'_{{\rm T},4}$ are dominant.
$G'_{{\rm T},1}$ decreases with $\gamma_0$, while 
the other $G_{{\rm T},n}'$ with $n=2,3,4$ are
almost independent of $\gamma_0$.
This indicates that SAS results from the behavior of $G'_{{\rm T},1}$.

On the other hand, the loss modulus $G''_{\rm T}$ consists of two terms as
\begin{eqnarray}
  G_{\rm T}''  & = & G_{{\rm T},1}'' + G_{{\rm T},2}''
\end{eqnarray}
with
\begin{align}
G_{{\rm T},1}''  & = 
   - \frac{1}{L^2}\sum_{i,j}\left \langle \left ( \frac{b_{ij,x}^{(1)}}{\gamma_0} Y_{ij} + X_{ij}\frac{b_{ij,y}^{(1)}}{\gamma_0} \right )  \Psi(R_{ij})  \right \rangle, \\
G_{{\rm T},2}''  & =  
  - \frac{1}{L^2}\sum_{i,j}\left \langle X_{ij}Y_{ij}\Psi'(R_{ij}) R_{ij}
   \frac{\boldsymbol R_{ij}\cdot  \boldsymbol b_{ij}^{(1)}}{\gamma_0 R_{ij}^2} \right \rangle.
\end{align}
In Fig. \ref{Gp_components}, we show $G''_{{\rm T},n}$ with $n=1$ and $2$ in the absorbing state for $\omega = 10^{-4}\tau_0^{-1}$ with $\phi=0.870$.
The result shows that $G''_{{\rm T},1}$ is dominant and almost independent of $\gamma_0$. 
$G''_{{\rm T},2}$ depends on $\gamma_0$, but it is much smaller than $G''_{{\rm T},1}$ for $\gamma_0<0.1$.

\begin{figure}[htbp]
  \includegraphics[width=0.9\linewidth]{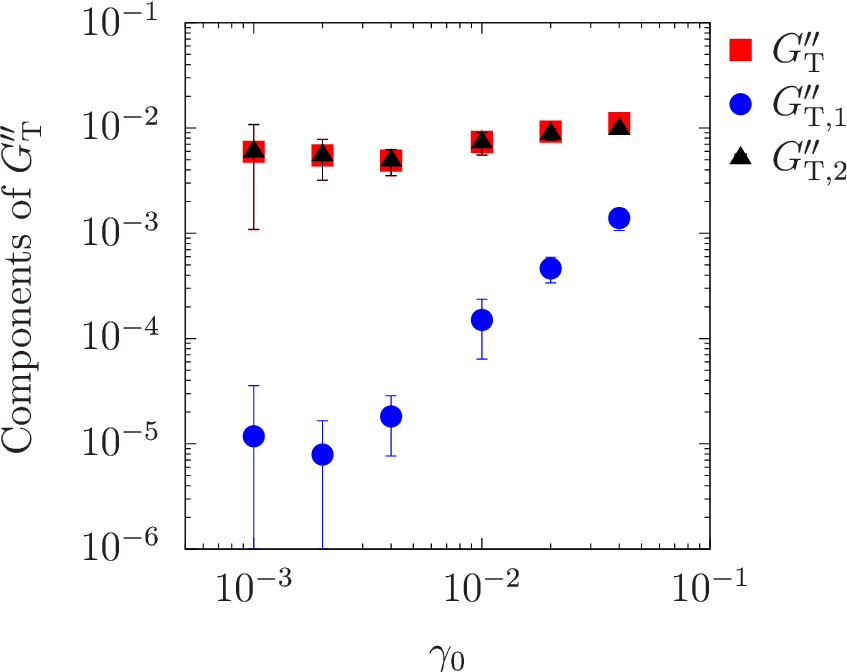}
  \caption{ $G'_{{\rm T},n}$ with $n=1$ and $2$ against $\gamma_0$ in the absorbing state for $\omega = 10^{-4}\tau_0^{-1}$ with $\phi=0.870$.
}
  \label{Gpp_components}
\end{figure}

\section{Non-linear viscoelastic moduli}
\label{Sec:HH}

\begin{figure}[htbp]
\includegraphics[width=0.9\linewidth]{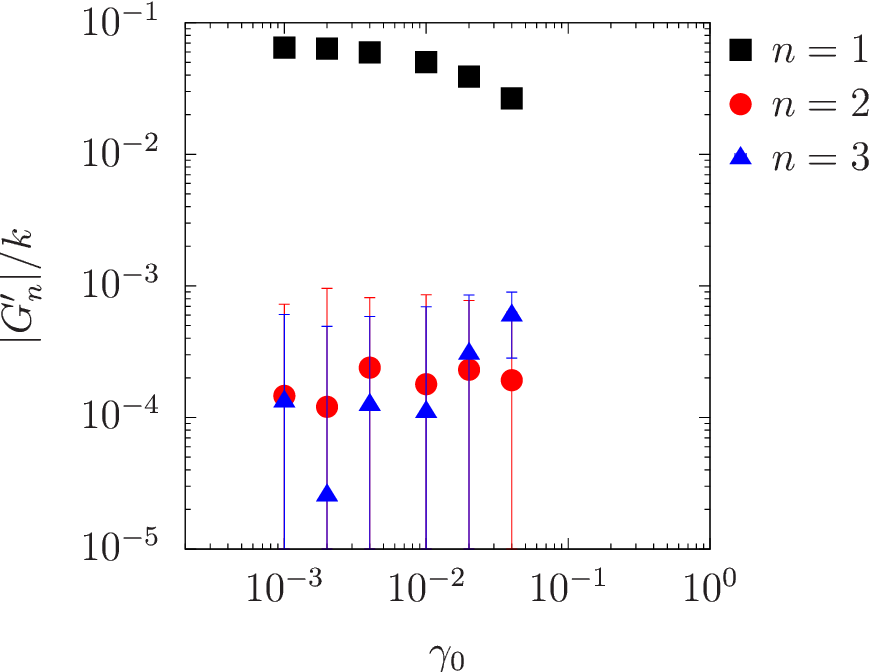}
  \caption{ $G_n'$ against $\gamma_0$ in the absorbing state for $\omega = 10^{-4}\tau_0^{-1}$ and $\phi=0.870$ with $n=1, 2$, and $3$.
}
\label{Gp_HH}
\end{figure}

\begin{figure}[htbp]
\includegraphics[width=0.9\linewidth]{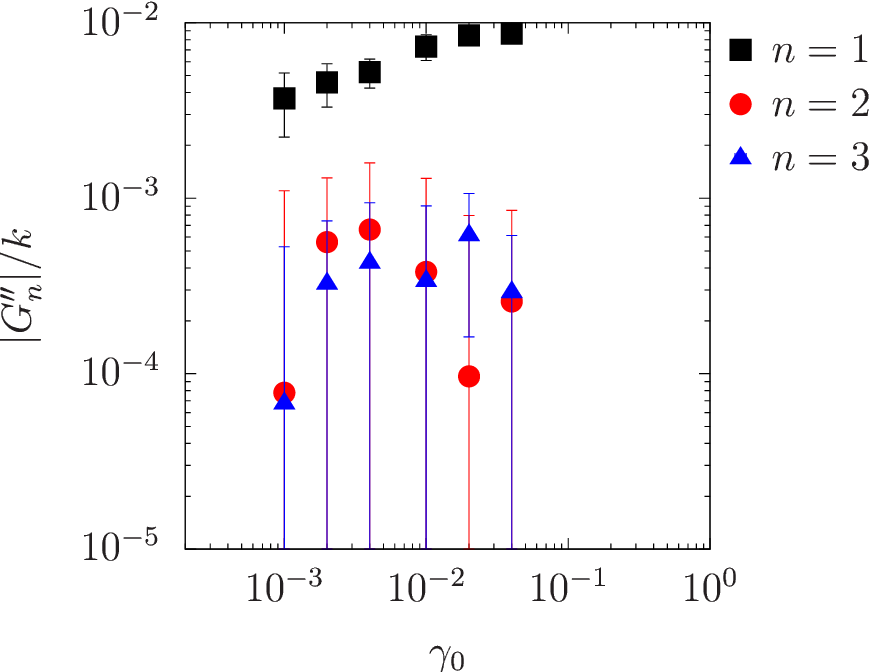}
  \caption{ $G_n''$ against $\gamma_0$ in the absorbing state for $\omega = 10^{-4}\tau_0^{-1}$ and $\phi=0.870$ with $n=1, 2$, and $3$.
}
\label{Gpp_HH}
\end{figure}

In this section, we examine the non-liner viscoelastic moduli in our system.
The nonlinear elastic response is generally characterized by nonlinear viscoelastic moduli $G'_n$ and $G''_n$ satisfying [41, 42]
\begin{equation}
  \sigma(t) = \gamma_0 \sum_{n=1} \{ G'_n \sin(n \omega t) + G''_n \cos(n \omega t)\}.
\end{equation}
The storage and loss moduli are, respectively, given by $G' = G'_1$ and $G'' = G''_1$. 
$G'_n$ and $G''_n$ for $n\ge 2$ represent higher harmonics.
In Figs. \ref{Gp_HH} and \ref{Gpp_HH}, we plot $G'_n$ and $G''_n$ in the absorbing state for $\omega = 10^{-4}\tau_0^{-1}$ and $\phi=0.870$ with $n=1, 2$, and $3$, respectively.
These figures indicate that the higher harmonics are negligible in our system.

%


\begin{thebibliography}{99}
\bibitem{Comment} We suppose $a^{(1)}$ is the largest because the mode proportional to $a^{(1)}$ is synchronized with the external oscillation $\sin \theta$.
\end{thebibliography}
\end{document}